\documentclass[aps,prd,onecolumn,showpacs,showkeys,amsmath,amssymb,nofootinbib]{revtex4}
\usepackage{graphicx}
\usepackage{dcolumn}
\usepackage{amsmath}
\usepackage{bm}
\usepackage{yhmath}
\usepackage{hyperref}
\usepackage{booktabs}
\usepackage{color}

\begin{document}

\def\Tr{{\rm Tr}}
\newcommand{\sla}{\not\!}

\allowdisplaybreaks

\title{Pseudoscalar Meson and Charmed Baryon Scattering Lengths}

\author{Zhan-Wei Liu}\email{liuzhanwei@pku.edu.cn}
\author{Shi-Lin Zhu}\email{zhusl@pku.edu.cn}
\affiliation{Department of Physics
and State Key Laboratory of Nuclear Physics and Technology,\\
Peking University, Beijing 100871, China }

\begin{abstract}
We have calculated the scattering lengths between the pseudoscalar
meson and charmed triplet, sextet, and excited sextet baryon to
the third order with the heavy baryon chiral perturbation theory.
The chiral expansion of some pion and eta channels converges well.
\end{abstract}



\maketitle

\pagenumbering{arabic}

\section{Introduction}\label{secIntr}
Up to now many heavy hadrons have been discovered experimentally.
Their inner structures and interactions attract much attention.
Some of them are speculated to be possible new hadron states
beyond the traditional quark model. For example, the newly
discovered $Z_b$ states are treated as the $\bar B^{(*)}B^*$
molecular states \cite{Adachi2011,Sun2011}. Many other molecular
states, such as those composed of $\Xi_c \Xi_c$, $D \bar D^*$, are
also proposed
\cite{Voloshin2004,Lee2011,Meguro2011,Wong2004,Close2004,Liu2008a}.
Whether there is attractive interaction between the particles is
the most important condition to form molecular states.

On the other hand, the hadron-hadron interaction may distort the
conventional quark model spectrum through the coupled channel
effects. For example, the bare charm-strange scalar meson lies
around 2.4 GeV according to the quark model calculation.
Experimentally the mass of the $D_{s0}(2317)$ state was measured
to be around 2.3 GeV. The attractive interaction between the $D$
meson and kaon is essential to lower its mass through the coupling
effect between the bare $c\bar s$ state and the $DK$ continuum
\cite{Dai2008}.

There has been lots of research work on the strong interactions of
the charmed or bottomed mesons, such as the lattice study,
calculations with the chiral perturbation theory and so on
\cite{Flynn2007,Ohki2008,Detmold2011,Gong2011,Wise1992,Geng2010,Liu2011a,Gamermann2007,Guo2009,Lutz2011}.
The charmed triplet ($B_{\bar 3}$), sextet ($B_6$) and excited
sextet ($B_6^*$) baryons are relatively stable particles. The pair
from the ground and corresponding excited sextet form a degenerate
doublet in the heavy quark spin symmetry limit. They interact with
other particles through the exchange of the pseudoscalar mesons in
the low energy effective field theory. It is very important to
study the strong interaction between the lightest pseudoscalar
meson ($\phi$) and charmed baryon.

A physical observable such as the scattering amplitude can be
expanded order by order with the explicit power counting in the
heavy baryon chiral perturbation theory (HB$\chi$PT). The
inclusion of the nonanalytic corrections resulting from the loop
diagrams would highly reduce the error of extraction in the
lattice study \cite{Detmold2012,Hall2011}, which is one of the
motivations of the present investigation.

In this work, we shall study the pseudoscalar meson and charmed
baryon scattering lengths to the third order with HB$\chi$PT. We
include the interaction of $B_6$ and $B_6^*$ explicitly for the
$\phi B_{\bar 3}$ scattering instead of absorbing their effects
into the low energy constants (LECs) at higher order since the
mass difference among the charmed baryons is small and the
couplings between them can not be neglected. The situation is
similar for the $\phi B_6$ and $\phi B_6^*$ cases. We express the
results as power series in $\epsilon=p/\Lambda_\chi$ with the
explicit power counting, where $p$ represents the mass and
momentum of pseudoscalar Goldstone bosons, the residual momentum
of charmed baryons, the mass difference among charmed baryons,
while $\Lambda_\chi$ is either the mass of charmed baryons or
$4\pi f$.

This paper is organized as follows: in Sec. \ref{secTMat}, we list
the HB$\chi$PT Lagrangians of the pseudoscalar mesons and charmed
baryons, with which we get the expressions of the $T$-matrices at
thresholds. In Sec. \ref{secLEC}, we estimate the LECs in the
Lagrangians. We present the numerical results and discussions in
Sec. \ref{secNum}.  Sec. \ref{secCon} is a short conclusion.

\section{The $T$-matrices at thresholds}\label{secTMat}

The average mass of the charmed triplet baryons $M_0(2408 {~\rm
MeV})$, which provides the base when we refer to the mass
difference in the following. The HB$\chi$PT Lagrangians at the
leading order read
\begin{eqnarray}
    {\mathcal L}_{\phi \phi}^{(2)}&=&f^2 \Tr\left( u_\mu u^\mu+\frac{\chi_+}{4} \right), \label{LPhiPhi} \\
    {\mathcal L}^{(1)}_{B\phi}&=&\frac12 \Tr[\bar B_{\bar 3} i v\cdot D B_{\bar 3}]
                            +\Tr[\bar B_{6} (i v\cdot D-\delta_2) B_{6}]
                            -\Tr[\bar B_{6}^* (i v\cdot D-\delta_3) B_{6}^*]\nonumber\\&&
                            +2 g_1 \Tr(\bar B_6 S\cdot u B_6)
                            +2 g_2 \Tr(\bar B_6 S\cdot u B_{\bar 3}+{\rm H.c.})
                            + g_3 \Tr(\bar B_{6 \mu}^* u^\mu B_{6}+{\rm H.c.})\nonumber\\&&
                            + g_4 \Tr(\bar B_{6 \mu}^* u^\mu B_{\bar 3}+{\rm H.c.})
                            +2 g_5 \Tr(\bar B_{6}^* S\cdot u B_{6}^*)
                            +2 g_6 \Tr(\bar B_{\bar 3} S\cdot u B_{\bar 3}), \label{LBPhiOne}
\end{eqnarray}
where $v_\mu$ is the velocity of a slowly moving baryon, $S_\mu$
is the spin matrix, $g_i$ is the coupling of the $BB\phi$-vertex,
and $\delta_i$ is the mass difference between the charmed baryons,
\begin{equation}
  \delta_1=M_{B_6^*}-M_{B_6}=67 {~\rm MeV}, \quad
  \delta_2=M_{B_6}-M_{B_{\bar3}}=127 {~\rm MeV}, \quad
  \delta_3=M_{B_6^*}-M_{B_{\bar3}}=194 {~\rm MeV}.
\end{equation}
The field notations are
\begin{equation}
\phi=\sqrt2\left(
\begin{array}{ccc}
\frac{\pi^0}{\sqrt2}+\frac{\eta}{\sqrt6}&\pi^+&K^+\\
\pi^-&-\frac{\pi^0}{\sqrt2}+\frac{\eta}{\sqrt6}&K^0\\
K^-&\overline{K}^0&-\frac{2}{\sqrt6}\eta
\end{array}\right),\quad
B_{\bar 3}=\left(
\begin{array}{ccc}
0&  \Lambda_c^+&    \Xi_c^+\\
-\Lambda_c^+&   0&  \Xi_c^0\\
-\Xi_c^+&-\Xi_c^0&0
\end{array}\right), \quad
B_6=\left( \begin{array}{ccc}
\Sigma_c^{++}&    \frac{\Sigma_c^+}{\sqrt 2} &    \frac{\Xi_c^{\prime+}}{\sqrt 2} \\
 \frac{\Sigma_c^+}{\sqrt 2} &  \Sigma_c^0    &\frac{\Xi_c^{\prime0}}{\sqrt 2} \\
\frac{\Xi_c^{\prime+}}{\sqrt 2} &   \frac{\Xi_c^{\prime0}}{\sqrt
2}  &\Omega_c^0
\end{array}\right),
\end{equation}
\begin{equation}
\Gamma_\mu = \frac{i}{2} [\xi^\dagger, \partial_\mu\xi],\quad
u_\mu=\frac{i}{2} \{\xi^\dagger, \partial_\mu \xi\},\quad \xi
=\exp(i \frac{\phi}{2f}), \quad \chi_\pm
=\xi^\dagger\chi\xi^\dagger\pm\xi\chi\xi,\quad
\chi=\mathrm{diag}(m_\pi^2,\, m_\pi^2,\, 2m_K^2-m_\pi^2),
\label{uxiDef}
\end{equation}
and the definition of $B_6^*$ is similar to that of $B_6$. The
covariant derivatives, $iD^\mu B_{ab}=i \partial^\mu
B_{ab}+\Gamma_a^{\mu~d} B_{db}+\Gamma_b^{\mu~d} B_{ad}$, will generate
the $BB\phi\phi$-vertexes.

The Lagrangian at $O(\epsilon^2)$ contains the counter terms and
the recoil terms. The counter terms are constructed on the basis
of the chiral and other symmetries and proportional to $\bar c_i,
c_i, \tilde c_i$ in Eq. (\ref{LBPhiTwo}). The recoil terms are
derived from the Lagrangians of the leading order and proportional
to $g_i^2$. We list the relevant terms below,
\begin{eqnarray}
  {\mathcal L}^{(2)}_{B\phi}&=&
    \bar c_0 \Tr[\bar B_{\bar 3} B_{\bar 3}]\Tr[\chi_+]
   +\bar c_1 \Tr[\bar B_{\bar 3} \tilde \chi_+ B_{\bar 3}]
   +\left(\bar c_2-\frac{2g_6^2+g_2^2}{4M_0}\right) \Tr[\bar B_{\bar 3} v\cdot u~ v\cdot u B_{\bar 3}]\nonumber\\&&
   +\left(\bar c_3-\frac{2g_6^2-g_2^2}{4M_0}\right) \bar B_{\bar 3}^{ab} v\cdot u_a^{~c}~ v\cdot u_b^{~d} B_{\bar 3,cd}\nonumber\\&&
   +c_0 \Tr[\bar B_6 B_6]\Tr[\chi_+]
   +c_1 \Tr[\bar B_6 \tilde \chi_+ B_6]
   +\left(c_2-\frac{2g_2^2+g_1^2}{4M_0}\right) \Tr[\bar B_6 v\cdot u~ v\cdot u B_6]\nonumber\\&&
   +\left(c_3+\frac{2g_2^2-g_1^2}{4M_0}\right) \bar B_6^{ab} v\cdot u_a^{~c}~ v\cdot u_b^{~d} B_{6,cd}
   +c_4 \Tr[\bar B_6 B_6] \Tr[ v\cdot u~ v\cdot u] \nonumber\\&&
   -\tilde c_0 \Tr[\bar B_6^* B_6^*]\Tr[\chi_+]
   -\tilde c_1 \Tr[\bar B_6^* \tilde \chi_+ B_6^*]
   -\left(\tilde c_2-\frac{g_5^2}{4M_0} \right)\Tr[\bar B_6^* v\cdot u~ v\cdot u B_6^*]\nonumber\\&&
   -\left(\tilde c_3-\frac{g_5^2}{4M_0} \right)\bar B_6^{*ab} v\cdot u_a^{~c}~ v\cdot u_b^{~d} B_{6,cd}^*
   -\tilde c_4 \Tr[\bar B_6^* B_6^*] \Tr[ v\cdot u~ v\cdot u],  \label{LBPhiTwo}
\end{eqnarray}
where the traceless $\tilde\chi_\pm$ are defined as:
$\tilde\chi_\pm=\chi_\pm-\frac13\Tr[\chi_\pm]$.

The Lagrangian at $O(\epsilon^3)$ also contains the recoil-term
part,
\begin{eqnarray}
  {\mathcal L}^{(3,r)}_{B\phi}&=&
   \frac{2g_6^2+g_2^2}{8M_0^2} \Tr[\bar B_{\bar 3} v\cdot u(iv\cdot D) v\cdot u B_{\bar 3}]
   +\frac{2g_6^2-g_2^2}{8M_0^2}\bar B_{\bar 3}^{ab} v\cdot u_a^{~c}(iv\cdot D) v\cdot u_b^{~d} B_{\bar 3,cd}
   +\frac{g_2^2}{8M_0^2}\Tr[\bar B_{\bar 3} v\cdot u~v\cdot u B_{\bar 3}]\delta_2\nonumber\\&&
   -\frac{g_2^2}{8M_0^2}\bar B_{\bar 3}^{ab} v\cdot u_a^{~c} v\cdot u_b^{~d} B_{\bar 3,cd} \delta_2
   +\frac{2g_2^2+g_1^2}{8M_0^2} \Tr[\bar B_6 v\cdot u (iv\cdot D) v\cdot u B_6]\nonumber\\&&
   -\frac{2g_2^2-g_1^2}{8M_0^2} \bar B_6^{ab} v\cdot u_a^{~c}(iv\cdot D) v\cdot u_b^{~d} B_{6,cd}
   +\frac{g_1^2}{8M_0^2} \Tr[\bar B_6 v\cdot u~v\cdot u B_6]\delta_2
   +\frac{g_1^2}{8M_0^2} \bar B_6^{ab} v\cdot u_a^{~c} v\cdot u_b^{~d} B_{6,cd}\delta_2\nonumber\\&&
   -\frac{g_5^2}{8M_0^2}\Tr[\bar B_6^* v\cdot u(iv\cdot D+\delta_3) v\cdot u B_6^*]
   -\frac{g_5^2}{8M_0^2}\bar B_6^{*ab} v\cdot u_a^{~c}(iv\cdot D+\delta_3) v\cdot u_b^{~d} B_{6,cd}^*. \label{LBPhiTriR}
\end{eqnarray}

We neglect the contributions from the finite counter-term part at
$O(\epsilon^3)$ as in Refs. \cite{Kaiser2001,Liu2007}, and cancel
the divergences of the loop diagrams with the following infinite
part
\begin{eqnarray}
  {\mathcal L}^{(3,c)}_{B\phi}&=&
    \frac3{4f^2} L \Tr\left[\bar B_{\bar 3}[v\cdot u, \tilde\chi_-] B_{\bar 3}\right]
   -(\frac{5\alpha g_2^2 \delta_2}{2f^2}   -\frac{5\beta g_4^2 \delta_3}{8f^2}  ) L \Tr[\bar B_{\bar 3} \tilde \chi_+ B_{\bar 3}]
   +(\frac{10\alpha g_2^2 \delta_2}{3f^2} - \frac{5\beta g_4^2  \delta_3}{6f^2})  L \Tr[\bar B_{\bar 3} v\cdot u~ v\cdot u B_{\bar 3}]\nonumber\\&&
   -(\frac{32\alpha g_2^2 \delta_2}{3f^2} - \frac{8\beta g_4^2  \delta_3}{3f^2} ) L  \bar B_{\bar 3}^{ab} v\cdot u_a^{~c}~ v\cdot u_b^{~d} B_{\bar 3,cd}\nonumber\\&&
   +\frac{3}{2f^2} L \Tr\left[\bar B_6 [v\cdot u, \tilde\chi_-] B_6\right]
   +(\frac{\alpha g_2^2 \delta_2}{f^2}  +\frac{7\beta g_3^2 \delta_1}{24f^2}  )L \Tr[\bar B_6 \tilde \chi_+ B_6]
   +(\frac{4\alpha g_2^2 \delta_2}{f^2}+\frac{5\beta g_3^2\delta_1}{2f^2})L \Tr[\bar B_6 v\cdot u~ v\cdot u B_6]\nonumber\\&&
   +(\frac{8\alpha g_2^2 \delta_2}{f^2}-\frac{\beta g_3^2\delta_1}{f^2})L \bar B_6^{ab} v\cdot u_a^{~c}~ v\cdot u_b^{~d} B_{6,cd}
   -(\frac{4\alpha g_2^2 \delta_2}{3f^2}+\frac{11\beta g_3^2\delta_1}{18f^2})L \Tr[\bar B_6 B_6] \Tr[ v\cdot u~ v\cdot u] \nonumber\\&&
   -\frac{3}{2f^2}L \Tr\left[\bar B_6^*[v\cdot u, \tilde\chi_-] B_6^*\right]
   +(\frac{7g_3^2\delta_1}{24f^2}+\frac{g_4^2 \delta_3}{4f^2} )L \Tr[\bar B_6^* \tilde \chi_+ B_6^*]
   +(\frac{5g_3^2\delta_1}{2f^2}+ \frac{g_4^2 \delta_3}{f^2})L\Tr[\bar B_6^* v\cdot u~ v\cdot u B_6^*]\nonumber\\&&
   -(\frac{g_3^2\delta_1}{f^2}-\frac{2 g_4^2 \delta_3}{f^2})L \bar B_6^{*ab} v\cdot u_a^{~c}~ v\cdot u_b^{~d} B_{6,cd}^*
   -(\frac{11g_3^2\delta_1}{18f^2}+\frac{g_4^2 \delta_3}{3f^2} )L \Tr[\bar B_6^* B_6^*] \Tr[ v\cdot u~ v\cdot u],  \label{LBPhiTriC}
\end{eqnarray}
where in the $D$ dimensional space-time,
\begin{equation}
    \alpha=S^2=-3/4-(D-4)/4,\quad
    \beta={P^{3/2}}{}_\mu^\mu=2+(D-4),\quad
    L=\frac{\lambda^{D-4}}{16\pi^2}\left\{\frac1{D-4}+\frac12(\gamma_E-1-\ln 4\pi)\right\},
\end{equation}
$P^{3/2}_{\mu\nu}$ is the projection operator for the
Rarita-Schwinger field, $\gamma_E= 0.5772157$ is the Euler
constant, $\lambda = 4\pi f$ is the energy scale.

The scattering length $a_{\phi B}$ is related to the threshold
$T$-matrix $T_{\phi B}$ by $T_{\phi B}=4\pi(1+m_\phi/m_B)a_{\phi
B}$. At the leading order, only the $BB\phi\phi$-vertex from the
contact terms in ${\mathcal L}^{(1)}_{B\phi}$ contributes to the
$T$-matrices at the threshold. At the second order, the
corresponding $BB\phi\phi$-vertex of ${\mathcal L}^{(2)}_{B\phi}$
contributes. At the third order, in addition to the contribution
of ${\mathcal L}^{(3,r/c)}_{B\phi}$, the $T$-matrices also receive
the contribution from the loop diagrams consisting of the vertices
in the leading order Lagrangian.

\begin{figure}[!htbp]
\centering \scalebox{1}{
  \begin{minipage}[c]{0.5\textwidth}
   \centering
   \scalebox{0.9}{\includegraphics{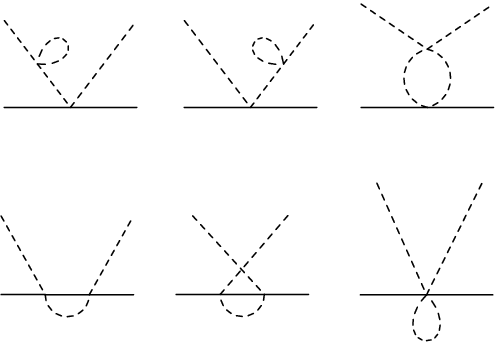}}\\
   I:
   no axial vertexes
   \end{minipage}
    \hfill
 \begin{minipage}[c]{0.5\textwidth}
   \centering
   \scalebox{0.9}{\includegraphics{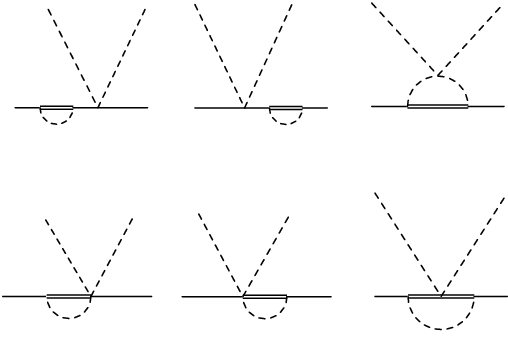}}\\
   II:
   two axial vertexes
 \end{minipage}
 }
\caption{Nonvanishing loop diagrams for the pseudoscalar meson and
charmed meson scattering lengths to $O(\epsilon^3)$ with
HB$\chi$PT. The dashed lines represent the pseudoscalar Goldstone
bosons. Both the thin solid lines and thick solid lines
represent charmed baryons. The internal thin solid lines represent
the charmed baryons in the same representation as the external
baryons while the internal thick solid lines represent all
possible charmed baryons. } \label{LoopDiag}
 \end{figure}

The nonvanishing loop diagrams are shown in Fig. \ref{LoopDiag}.
Since there is no vertex like $B_{\bar 3}B_6\phi\phi$ at the
leading order, the charmed baryons in different representations do
not appear in the diagrams (I) as intermediate states. But they
appear in the diagrams (II) through the axial coupling.

We calculate the loop diagrams with dimensional regularization and
the modified minimal subtraction scheme. We use the LECs in Eq.
(\ref{LBPhiTriC}) to cancel the divergence. At last we express the
$T$-matrices in terms of $f_\phi$ rather than $f$ with the help of
their relation in Refs. \cite{Gasser1985,Liu2011}.

The isospin symmetry is explicitly kept throughout our
calculation. So we only list the 49 isospin-independent
$T$-matrices in the following subsections.

\subsection{$\phi B_{\bar 3}$ scattering}

We list the $T$-matrices for the pseudoscalar meson and $\bar3$
charmed baryon scattering, separate them order by order with
braces, and distinguish between the tree and loop diagram
contribution at $O(\epsilon^3)$ with square brackets,
\begin{eqnarray}
T^{(1)}_{ \pi\Lambda_c}&=&\left\{0\right\}+\left\{-\frac{m_\pi^2 (24 \bar c_0+4 \bar c_1-3 \bar C_2-3 \bar C_3)}{3 f_\pi^2}\right\}+\left\{ [\frac{m_\pi^2 \delta_2 (\bar d_2+\bar d_3)}{f_\pi^2 M_0^2}] +[-\frac{1}{16} V\left(m_K^2,-m_\pi\right)
-\frac{1}{16} V\left(m_K^2,m_\pi\right)
\right.\nonumber\\&&\left.
-\frac{4}{9} m_\pi^2 Y_1(m_\eta)+2 m_\pi^2 W_1(m_\pi)] \right\}
,\nonumber\\
T^{(1/2)}_{ \pi\Xi_c}&=&\left\{\frac{m_\pi}{f_\pi^2}\right\}+\left\{-\frac{m_\pi^2 (48 \bar c_0-4 \bar c_1-3 \bar C_2)}{6 f_\pi^2}\right\}+\left\{ [\frac{m_\pi^2 (16 \bar d_1 m_\pi+\bar d_2 \delta_2)}{2 f_\pi^2 M_0^2}] +[\frac{1}{32} V\left(m_K^2,-m_\pi\right)-\frac{3}{32} V\left(m_K^2,m_\pi\right)
\right.\nonumber\\&&\left.
-\frac{1}{4} V\left(m_\pi^2,m_\pi\right)-\frac{1}{2} m_\pi^2 W_1(m_\eta)-\frac{1}{9} m_\pi^2 Y_1(m_\eta)+\frac{1}{2} m_\pi^2 W_1(m_\pi)+m_\pi^2 Y_1(m_\pi)] \right\}
,\nonumber\\
T^{(3/2)}_{ \pi\Xi_c}&=&\left\{-\frac{m_\pi}{2 f_\pi^2}\right\}+\left\{-\frac{m_\pi^2 (48 \bar c_0-4 \bar c_1-3 \bar C_2)}{6 f_\pi^2}\right\}+\left\{ [-\frac{m_\pi^2 (8 \bar d_1 m_\pi-\bar d_2 \delta_2)}{2 f_\pi^2 M_0^2}] +[-\frac{1}{16} V\left(m_K^2,-m_\pi\right)
\right.\nonumber\\&&\left.
-\frac{3}{16} V\left(m_\pi^2,-m_\pi\right)-\frac{1}{16} V\left(m_\pi^2,m_\pi\right)-\frac{1}{2} m_\pi^2 W_1(m_\eta)-\frac{1}{9} m_\pi^2 Y_1(m_\eta)+\frac{1}{2} m_\pi^2 W_1(m_\pi)+m_\pi^2 Y_1(m_\pi)] \right\}
,\nonumber\\
T^{(1/2)}_{ K\Lambda_c}&=&\left\{-\frac{m_K}{2 f_K^2}\right\}+\left\{-\frac{m_K^2 (48 \bar c_0-4 \bar c_1-3 \bar C_2)}{6 f_K^2}\right\}+\left\{ [-\frac{m_K^2 (8 \bar d_1 m_K-\bar d_2 \delta_2)}{2 f_K^2 M_0^2}] +[-\frac{1}{16} V\left(m_K^2,-m_K\right)
\right.\nonumber\\&&\left.
-\frac{1}{16} V\left(m_K^2,m_K\right)-\frac{3}{32} V\left(m_\eta^2,-m_K\right)-\frac{3}{32} V\left(m_\pi^2,-m_K\right)+\frac{8}{9} m_K^2 Y_1(m_\eta)] \right\}
,\nonumber\\
T^{(0)}_{ K\Xi_c}&=&\left\{\frac{m_K}{f_K^2}\right\}+\left\{-\frac{m_K^2 (24 \bar c_0-8 \bar c_1+3 \bar C_3)}{3 f_K^2}\right\}+\left\{ [\frac{m_K^2 (8 \bar d_1 m_K-\bar d_3 \delta_2)}{f_K^2 M_0^2}] +[-\frac{1}{16} V\left(m_K^2,-m_K\right)-\frac{1}{4} V\left(m_K^2,m_K\right)
\right.\nonumber\\&&\left.
-\frac{3}{16} V\left(m_\eta^2,m_K\right)+m_K^2 W_1(m_\eta)+\frac{2}{9} m_K^2 Y_1(m_\eta)+2 P_1-3 U_1] \right\}
,\nonumber\\
T^{(1)}_{ K\Xi_c}&=&\left\{0\right\}+\left\{-\frac{m_K^2 (24 \bar c_0+4 \bar c_1-3 \bar C_2-3 \bar C_3)}{3 f_K^2}\right\}+\left\{ [\frac{m_K^2 \delta_2 (\bar d_2+\bar d_3)}{f_K^2 M_0^2}] +[-\frac{1}{16} V\left(m_K^2,-m_K\right)-\frac{1}{16} V\left(m_\pi^2,m_K\right)
\right.\nonumber\\&&\left.
+m_K^2 W_1(m_\eta)+\frac{2}{9} m_K^2 Y_1(m_\eta)-\frac{2 P_1}{3}+U_1] \right\}
,\nonumber\\
T^{(0)}_{ \eta\Lambda_c}&=&\left\{0\right\}+\left\{-\frac{24 \bar c_0 m_\eta^2-8 \bar c_1 m_\eta^2+4 \bar c_1 m_\pi^2-\bar C_2 m_\eta^2+\bar C_3 m_\eta^2}{3 f_\eta^2}\right\}+\left\{ [\frac{m_\eta^2 \delta_2 (\bar d_2-\bar d_3)}{3 f_\eta^2 M_0^2}] +[-\frac{3}{16} V\left(m_K^2,-m_\eta\right)
\right.\nonumber\\&&\left.
-\frac{3}{16} V\left(m_K^2,m_\eta\right)-\frac{64}{27} m_K^2 Y_1(m_\eta)+\frac{4}{3} m_K^2 W_1(m_K)+\frac{8}{3} m_K^2 Y_1(m_K)+\frac{28}{27} m_\pi^2 Y_1(m_\eta)-2 m_\pi^2 W_1(m_\pi)] \right\}
,\nonumber\\
T^{(1/2)}_{ \eta\Xi_c}&=&\left\{0\right\}+\left\{-\frac{48 \bar c_0 m_\eta^2+8 \bar c_1 m_\eta^2-4 \bar c_1 m_\pi^2-5 \bar C_2 m_\eta^2-4 \bar C_3 m_\eta^2}{6 f_\eta^2}\right\}+\left\{ [\frac{m_\eta^2 \delta_2 (5 \bar d_2+4 \bar d_3)}{6 f_\eta^2 M_0^2}]
+[-\frac{3}{32} V\left(m_K^2,-m_\eta\right)
\right.\nonumber\\&&\left.
-\frac{3}{32} V\left(m_K^2,m_\eta\right)-\frac{8}{3} m_K^2 W_1(m_\eta)-\frac{16}{27} m_K^2 Y_1(m_\eta)+\frac{10}{3} m_K^2 W_1(m_K)+\frac{4}{3} m_K^2 Y_1(m_K)+\frac{7}{6} m_\pi^2 W_1(m_\eta)
\right.\nonumber\\&&\left.
+\frac{7}{27} m_\pi^2 Y_1(m_\eta)-\frac{1}{2} m_\pi^2 W_1(m_\pi)-m_\pi^2 Y_1(m_\pi)] \right\}\label{TphiB3}
\end{eqnarray}
where the superscript $I$ in $T^{I}_{\phi B}$ refers to the
isospin of the channel, the functions $P$, $U$, $V$, $W$, and $Y$
are listed in Appendix \ref{secFunc}, and some combination
coefficients are defined as
\begin{equation}
  \bar C_2=\bar c_2-\frac{2g_6^2+g_2^2}{4M_0},\quad
  \bar C_3=\bar c_3-\frac{2g_6^2-g_2^2}{4M_0},\quad
  \bar d_1=\frac{2g_6^2+g_2^2}{64},\quad
  \bar d_2=\frac{g_2^2}{8},\quad
  \bar d_3=-\frac{g_2^2}{8},\quad
\end{equation}
We have used the Gell-Mann-Okubo mass relation
$m_\eta^2=(4m_K^2-m_\pi^2)/3$ to make the expression more concise.

Besides the eight $T$-matrices listed above, the other three
isospin-independent ones can be written in terms of those in Eq.
(\ref{TphiB3}) by crossing symmetry,
\begin{equation}
  T^{(1/2)}_{ \bar K\Lambda_c}=\left[T^{(1/2)}_{ K\Lambda_c}\right]_{m_K\rightarrow -m_K},\quad
  T^{(1)}_{ \bar K\Xi_c}=\frac12\left[T^{(1)}_{K\Xi_c}+T^{(0)}_{K\Xi_c}\right]_{m_K\rightarrow -m_K},\quad
  T^{(0)}_{ \bar K\Xi_c}=\frac12\left[3T^{(1)}_{K\Xi_c}-T^{(0)}_{K\Xi_c}\right]_{m_K\rightarrow -m_K}.\quad
\end{equation}

\subsection{$\phi B_6$ scattering}

There are 19 isospin independent $T$-matrices for the pseudoscalar
meson and charmed sextet baryon scattering.
\begin{eqnarray}
T^{(1)}_{ \pi\Omega_c}&=&\left\{0\right\}+\left\{-\frac{m_\pi^2
(12 c_0-4 c_1-3 c_4)}{3 f_\pi^2}\right\}+\left\{ [0]
+[-\frac{1}{8} V\left(m_K^2,-m_\pi\right)-\frac{1}{8}
V\left(m_K^2,m_\pi\right)-\frac{4}{9} m_\pi^2 W_2(m_\eta)]
\right\},
\nonumber\\
T^{(1/2)}_{
\pi\Xi'_c}&=&\left\{\frac{m_\pi}{f_\pi^2}\right\}+\left\{-\frac{m_\pi^2
(48 c_0-4 c_1-3 C_2-12 c_4)}{12 f_\pi^2}\right\}+\left\{
[\frac{m_\pi^2 (16 d_1 m_\pi+d_2 \delta_2)}{4 f_\pi^2 M_0^2}]
+[-\frac{3}{32} V\left(m_K^2,-m_\pi\right)
\right.\nonumber\\&&\left. -\frac{7}{32}
V\left(m_K^2,m_\pi\right)-\frac{1}{4}
V\left(m_\pi^2,m_\pi\right)-\frac{1}{36} m_\pi^2
W_2(m_\eta)-\frac{1}{2} m_\pi^2 Y_2(m_\eta)+\frac{1}{4} m_\pi^2
W_2(m_\pi)+\frac{1}{2} m_\pi^2 Y_2(m_\pi)] \right\},
\nonumber\\
T^{(3/2)}_{ \pi\Xi'_c}&=&\left\{-\frac{m_\pi}{2
f_\pi^2}\right\}+\left\{-\frac{m_\pi^2 (48 c_0-4 c_1-3 C_2-12
c_4)}{12 f_\pi^2}\right\}+\left\{ [-\frac{m_\pi^2 (8 d_1 m_\pi-d_2
\delta_2)}{4 f_\pi^2 M_0^2}] +[-\frac{3}{16}
V\left(m_K^2,-m_\pi\right) \right.\nonumber\\&&\left. -\frac{1}{8}
V\left(m_K^2,m_\pi\right)-\frac{3}{16}
V\left(m_\pi^2,-m_\pi\right)-\frac{1}{16}
V\left(m_\pi^2,m_\pi\right)-\frac{1}{36} m_\pi^2
W_2(m_\eta)-\frac{1}{2} m_\pi^2 Y_2(m_\eta)+\frac{1}{4} m_\pi^2
W_2(m_\pi) \right.\nonumber\\&&\left. +\frac{1}{2} m_\pi^2
Y_2(m_\pi)] \right\},
\nonumber\\
T^{(0)}_{ \pi\Sigma_c}&=&\left\{\frac{2
m_\pi}{f_\pi^2}\right\}+\left\{-\frac{m_\pi^2 (24 c_0+4 c_1-3
C_2+9 C_3-6 c_4)}{6 f_\pi^2}\right\}+\left\{ [\frac{m_\pi^2 (16
d_1 m_\pi+d_2 \delta_2-3 d_3 \delta_2)}{2 f_\pi^2 M_0^2}]
\right.\nonumber\\&&\left. +[\frac{1}{16}
V\left(m_K^2,-m_\pi\right)-\frac{3}{16}
V\left(m_K^2,m_\pi\right)-\frac{1}{2}
V\left(m_\pi^2,-m_\pi\right)-V\left(m_\pi^2,m_\pi\right)-\frac{1}{9}
m_\pi^2 W_2(m_\eta)+4 m_\pi^2 W_2(m_\pi)
\right.\nonumber\\&&\left. -6 m_\pi^2 Y_2(m_\pi)] \right\},
\nonumber\\
T^{(1)}_{
\pi\Sigma_c}&=&\left\{\frac{m_\pi}{f_\pi^2}\right\}+\left\{-\frac{m_\pi^2
(24 c_0+4 c_1-3 C_2-6 C_3-6 c_4)}{6 f_\pi^2}\right\}+\left\{
[\frac{m_\pi^2 (8 d_1 m_\pi+d_2 \delta_2+2 d_3 \delta_2)}{2
f_\pi^2 M_0^2}] +[-\frac{1}{8} V\left(m_K^2,m_\pi\right)
\right.\nonumber\\&&\left. -\frac{1}{4}
V\left(m_\pi^2,m_\pi\right)-\frac{1}{9} m_\pi^2
W_2(m_\eta)-m_\pi^2 W_2(m_\pi)+4 m_\pi^2 Y_2(m_\pi)] \right\},
\nonumber\\
T^{(2)}_{
\pi\Sigma_c}&=&\left\{-\frac{m_\pi}{f_\pi^2}\right\}+\left\{-\frac{m_\pi^2
(24 c_0+4 c_1-3 C_2-6 c_4)}{6 f_\pi^2}\right\}+\left\{
[-\frac{m_\pi^2 (8 d_1 m_\pi-d_2 \delta_2)}{2 f_\pi^2 M_0^2}]
+[-\frac{1}{8} V\left(m_K^2,-m_\pi\right)
\right.\nonumber\\&&\left. -\frac{1}{2}
V\left(m_\pi^2,-m_\pi\right)-\frac{1}{4}
V\left(m_\pi^2,m_\pi\right)-\frac{1}{9} m_\pi^2
W_2(m_\eta)+m_\pi^2 W_2(m_\pi)] \right\},
\nonumber\\
T^{(1/2)}_{
K\Omega_c}&=&\left\{\frac{m_K}{f_K^2}\right\}+\left\{-\frac{m_K^2
(24 c_0+4 c_1-3 C_2-6 c_4)}{6 f_K^2}\right\}+\left\{ [\frac{m_K^2
(8 d_1 m_K+d_2 \delta_2)}{2 f_K^2 M_0^2}] +[-\frac{1}{4}
V\left(m_K^2,-m_K\right) \right.\nonumber\\&&\left. -\frac{1}{4}
V\left(m_K^2,m_K\right)-\frac{3}{16}
V\left(m_\eta^2,m_K\right)-\frac{3}{16}
V\left(m_\pi^2,m_K\right)+\frac{8}{9} m_K^2 W_2(m_\eta)] \right\},
\nonumber\\
T^{(0)}_{
K\Xi'_c}&=&\left\{\frac{m_K}{f_K^2}\right\}+\left\{-\frac{m_K^2
(24 c_0-8 c_1+3 C_3-6 c_4)}{6 f_K^2}\right\}+\left\{ [\frac{m_K^2
(8 d_1 m_K-d_3 \delta_2)}{2 f_K^2 M_0^2}] +[-\frac{1}{16}
V\left(m_K^2,-m_K\right) \right.\nonumber\\&&\left. -\frac{1}{4}
V\left(m_K^2,m_K\right)+\frac{3}{16}
V\left(m_\eta^2,-m_K\right)-\frac{3}{16}
V\left(m_\pi^2,-m_K\right)-\frac{3}{16}
V\left(m_\pi^2,m_K\right)+\frac{1}{18} m_K^2 W_2(m_\eta)
\right.\nonumber\\&&\left. +m_K^2 Y_2(m_\eta)-3 P_2+\frac{U_2}{2}]
\right\},
\nonumber\\
T^{(1)}_{ K\Xi'_c}&=&\left\{0\right\}+\left\{-\frac{m_K^2 (24
c_0+4 c_1-3 C_2-3 C_3-6 c_4)}{6 f_K^2}\right\}+\left\{
[\frac{m_K^2 \delta_2 (d_2+d_3)}{2 f_K^2 M_0^2}] +[-\frac{1}{16}
V\left(m_K^2,-m_K\right) \right.\nonumber\\&&\left. -\frac{3}{16}
V\left(m_\eta^2,-m_K\right)-\frac{3}{16}
V\left(m_\eta^2,m_K\right)-\frac{1}{16}
V\left(m_\pi^2,-m_K\right)-\frac{1}{8}
V\left(m_\pi^2,m_K\right)+\frac{1}{18} m_K^2 W_2(m_\eta)
\right.\nonumber\\&&\left. +m_K^2 Y_2(m_\eta)+P_2-\frac{U_2}{6}]
\right\},
\nonumber\\
T^{(1/2)}_{ K\Sigma_c}&=&\left\{\frac{m_K}{2
f_K^2}\right\}+\left\{-\frac{m_K^2 (48 c_0-28 c_1+3 C_2-12
c_4)}{12 f_K^2}\right\}+\left\{ [\frac{m_K^2 (8 d_1 m_K-d_2
\delta_2)}{4 f_K^2 M_0^2}] +[\frac{3}{16} V\left(m_K^2,-m_K\right)
\right.\nonumber\\&&\left. -\frac{1}{16}
V\left(m_K^2,m_K\right)+\frac{3}{32}
V\left(m_\eta^2,-m_K\right)-\frac{5}{32}
V\left(m_\pi^2,-m_K\right)+\frac{2}{9} m_K^2 W_2(m_\eta)-\frac{4
U_2}{3}] \right\},
\nonumber\\
T^{(3/2)}_{
K\Sigma_c}&=&\left\{-\frac{m_K}{f_K^2}\right\}+\left\{-\frac{m_K^2
(24 c_0+4 c_1-3 C_2-6 c_4)}{6 f_K^2}\right\}+\left\{ [-\frac{m_K^2
(8 d_1 m_K-d_2 \delta_2)}{2 f_K^2 M_0^2}] +[-\frac{3}{8}
V\left(m_K^2,-m_K\right) \right.\nonumber\\&&\left. -\frac{1}{4}
V\left(m_K^2,m_K\right)-\frac{3}{16}
V\left(m_\eta^2,-m_K\right)-\frac{1}{16}
V\left(m_\pi^2,-m_K\right)+\frac{2}{9} m_K^2 W_2(m_\eta)+\frac{2
U_2}{3}] \right\},
\nonumber\\
T^{(0)}_{ \eta\Omega_c}&=&\left\{0\right\}+\left\{-\frac{12 c_0
m_\eta^2+8 c_1 m_\eta^2-4 c_1 m_\pi^2-2 C_2 m_\eta^2-2 C_3
m_\eta^2-3 c_4 m_\eta^2}{3 f_\eta^2}\right\}+\left\{ [\frac{2
m_\eta^2 \delta_2 (d_2+d_3)}{3 f_\eta^2 M_0^2}] +[
\right.\nonumber\\&&\left. -\frac{3}{8}
V\left(m_K^2,-m_\eta\right)-\frac{3}{8}
V\left(m_K^2,m_\eta\right)-\frac{64}{27} m_K^2
W_2(m_\eta)+\frac{4}{3} m_K^2 W_2(m_K)+\frac{8}{3} m_K^2
Y_2(m_K)+\frac{28}{27} m_\pi^2 W_2(m_\eta)] \right\},
\nonumber\\
T^{(1/2)}_{ \eta\Xi'_c}&=&\left\{0\right\}+\left\{-\frac{48 c_0
m_\eta^2+8 c_1 m_\eta^2-4 c_1 m_\pi^2-5 C_2 m_\eta^2+4 C_3
m_\eta^2-12 c_4 m_\eta^2}{12 f_\eta^2}\right\}+\left\{
[\frac{m_\eta^2 \delta_2 (5 d_2-4 d_3)}{12 f_\eta^2 M_0^2}] +[
\right.\nonumber\\&&\left. -\frac{15}{32}
V\left(m_K^2,-m_\eta\right)-\frac{15}{32}
V\left(m_K^2,m_\eta\right)-\frac{4}{27} m_K^2
W_2(m_\eta)-\frac{8}{3} m_K^2 Y_2(m_\eta)+\frac{5}{3} m_K^2
W_2(m_K)+\frac{2}{3} m_K^2 Y_2(m_K) \right.\nonumber\\&&\left.
+\frac{7}{108} m_\pi^2 W_2(m_\eta)+\frac{7}{6} m_\pi^2
Y_2(m_\eta)-\frac{1}{4} m_\pi^2 W_2(m_\pi)-\frac{1}{2} m_\pi^2
Y_2(m_\pi)] \right\},
\nonumber\\
T^{(1)}_{ \eta\Sigma_c}&=&\left\{0\right\}+\left\{-\frac{24 c_0
m_\eta^2-8 c_1 m_\eta^2+4 c_1 m_\pi^2-C_2 m_\eta^2-C_3 m_\eta^2-6
c_4 m_\eta^2}{6 f_\eta^2}\right\}+\left\{ [\frac{m_\eta^2 \delta_2
(d_2+d_3)}{6 f_\eta^2 M_0^2}] +[ \right.\nonumber\\&&\left.
-\frac{3}{16} V\left(m_K^2,-m_\eta\right)-\frac{3}{16}
V\left(m_K^2,m_\eta\right)-\frac{16}{27} m_K^2
W_2(m_\eta)+\frac{2}{3} m_K^2 W_2(m_K)+\frac{4}{3} m_K^2
Y_2(m_K)+\frac{7}{27} m_\pi^2 W_2(m_\eta)
\right.\nonumber\\&&\left. -\frac{2}{3} m_\pi^2
W_2(m_\pi)-\frac{2}{3} m_\pi^2 Y_2(m_\pi)] \right\},
\label{TphiB6}
\end{eqnarray}
where
\begin{equation}
  C_2=c_2-\frac{2g_2^2+g_1^2}{4M_0},\quad
  C_3=c_3+\frac{2g_2^2-g_1^2}{4M_0},\quad
  d_1=\frac{g_1^2+2g_2^2}{64},\quad
  d_2=\frac{g_1^2+g_2^2}{4} ,\quad
  d_3=\frac{g_1^2-g_2^2}{4}.
\end{equation}

Moreover, with crossing symmetry we get
\begin{eqnarray}
  &&T^{(3/2)}_{ \bar K\Sigma_c}=\frac13\left[T^{(3/2)}_{K\Sigma_c}+2T^{(1/2)}_{K\Sigma_c}\right]_{m_K\rightarrow -m_K},\quad
  T^{(1/2)}_{ \bar K\Sigma_c}=\frac13\left[4T^{(3/2)}_{K\Sigma_c}-T^{(1/2)}_{K\Sigma_c}\right]_{m_K\rightarrow -m_K},\quad
  T^{(1/2)}_{ \bar K\Omega_c}=\left[T^{(1/2)}_{ K\Omega_c}\right]_{m_K\rightarrow -m_K},\nonumber\\&&
  T^{(1)}_{ \bar K\Xi^\prime_c}=\frac12\left[T^{(1)}_{K\Xi^\prime_c}+T^{(0)}_{K\Xi^\prime_c}\right]_{m_K\rightarrow -m_K},\quad
  T^{(0)}_{ \bar K\Xi^\prime_c}=\frac12\left[3T^{(1)}_{K\Xi^\prime_c}-T^{(0)}_{K\Xi^\prime_c}\right]_{m_K\rightarrow -m_K}.\quad
  \label{TphiB6c}
\end{eqnarray}

\subsection{$\phi B_6^*$ scattering}

There are also 19 independent $T$-matrices for the scattering
between pseudoscalar meson and the excited charmed sextet baryon.
Even including the loop correction, we notice that $T^{(I)}_{\phi
B_6^*}$ is the same as $T^{(I)}_{\phi B_6}$ in the heavy baryon
symmetry limit. Taking into account of the heavy baryon symmetry
breaking effect, we obtain $T^{(I)}_{\phi B_6^*}$ after making the
following replacements in the expressions of the corresponding
$T^{(I)}_{\phi B_6}$ in Eqs (\ref{TphiB6},\ref{TphiB6c})
\begin{eqnarray}
  &&c_1\to \tilde c_1, \quad
  C_2\to \tilde c_2-\frac{g_5^2}{4M_0} , \quad
  C_3\to \tilde c_3-\frac{g_5^2}{4M_0}, \quad
  c_4\to \tilde c_4, \quad
  d_1\to \frac{g_5^2}{64}, \quad
  d_2\to \frac{g_5^2}{4}, \quad
  d_3\to \frac{g_5^2}{4}, \nonumber\\&&
  P_2\to P_3, \quad
  U_2\to U_3, \quad
  W_2(m_X)\to W_3(m_X), \quad
  Y_2(m_X)\to Y_3(m_X).
\end{eqnarray}

We have listed the $T$-matrices of the pseudoscalar meson and
charmed baryon scattering in the above three subsections. We have
assumed the SU(3) flavor symmetry and taken the SU(3) breaking
effect into account perturbatively. One can also study the
scattering of $\pi B_{\bar3}$, $\pi B_{6}$, and $\pi B_{6}^*$ with
SU(2) flavor symmetry. We can construct the relevant Lagrangians
with exact SU(2) chiral symmetry from the beginning.

Alternatively, we can extract SU(2) Lagrangians from the SU(3) Lagrangians in
Eqs. (\ref{LPhiPhi}), (\ref{LBPhiOne}), (\ref{LBPhiTwo}),
(\ref{LBPhiTriR}), and (\ref{LBPhiTriC}). Now the coupling
constants $g_i$ and other LECs are different from those in the
SU(3) case. Then we can obtain the SU(2) $T$-matrices from the
SU(3) ones. More specially, we may drop the terms proportional to
$V(m_K^2,\pm m_\pi)$, $Y_i(m_\eta)$, and $W_i(m_\eta)$ in the
SU(3) $T_{\pi B}$, and replace $g_1$ and other LECs with new
independent ones $g_{1\Sigma_c}$, $g_{1\Xi^\prime_c}$,
$g_{1\Omega_c}$ etc. After that we get the SU(2) $T_{\pi B}$. The
contributions of the dropped terms actually are absorbed by the
redefined LECs at $O(\epsilon^3)$.

Unfortunately, the investigation of the $\pi B$ scattering with
SU(2) chiral perturbation theory introduces more independent LECs.
Especially the SU(2) LECs at $O(\epsilon^3)$ can not be neglected.
We do not include the contribution from the kaon and eta
explicitly, which contribute to $O(\epsilon^3)$ LECs here. In the
following we will concentrate on the SU(3) case only.

\section{Low Energy Constants}\label{secLEC}

Similar to the nucleon case \cite{Bernard2006,procura2007}, the
chiral correction to the charmed baryon axial-vector coupling
would also be $O(\epsilon^2)$, which contributes to the
$T$-matrices at $O(\epsilon^4)$ or higher order, thus can be
neglected. Using $|g_2|=0.60$ and $|g_4|=1.0$ obtained by fitting
the decay widths of $\Sigma_c$ and $\Sigma_c^*$ \cite{Meguro2011},
$|g_1|=\sqrt{8/3}|g_2|$ with the quark model symmetry, and
$|g_3|=\sqrt3/2 |g_1|,~|g_5|=3/2 |g_1|,~|g_6|=0$ with heavy quark
spin symmetry, we have
\begin{equation}
  |g_1|=0.98,\quad
  |g_2|=0.60,\quad
  |g_3|=0.85,\quad
  |g_4|=1.0,\quad
  |g_5|=1.5,\quad
  |g_6|=0.
\end{equation}
We also need
\cite{PDG2006,Liu2007a,Escribano2005}
\begin{equation}
  m_\pi=140~{\rm MeV}, \quad
  m_K=494~{\rm MeV}, \quad
  f_\pi=92~{\rm MeV},\quad
  f_K=113~{\rm MeV},\quad
  f_\eta=1.2f_K.
\end{equation}

Since there are no available experimental data to extract the low
energy constants at $O(\epsilon^2)$, we utilize the crude SU(4)
flavor symmetry to make a rough estimate of some of these LECs in
Appendix \ref{secSUFour},
\begin{eqnarray}
&&\bar c_0=-0.32{~\rm GeV}^{-1},\quad
  \bar c_1=-0.52{~\rm GeV}^{-1},\quad
  \bar c_2=-1.78{~\rm GeV}^{-1}+\frac13\frac{\alpha'}{4\pi f},\quad
  \bar c_3=-0.03{~\rm GeV}^{-1}-\frac13\frac{\alpha'}{4\pi f},\nonumber\\&&
  c_0=-0.61{~\rm GeV}^{-1},\quad
  c_1=-0.98{~\rm GeV}^{-1},\quad
  c_2=-2.07{~\rm GeV}^{-1}-2\frac{\alpha'}{4\pi f},\quad
  c_3=-0.84{~\rm GeV}^{-1},\quad
  c_4=\frac{\alpha'}{4\pi f}.
\end{eqnarray}
We would assume $\tilde c_i=c_i$ with the heavy quark spin
symmetry in the numerical calculation. As for the dimensionless
LEC $\alpha'$, we will take it to be in the natural range of
[-1,1] as in Ref. \cite{Guo2009}.

\section{Numerical results and discussions}\label{secNum}

We list the $T$-matrices order by order for the pseudoscalar meson
and charmed baryon scattering in Tables \ref{TabTPhiB3},
\ref{TabTPhiB6}, \ref{TabTPhiB6s}. The positive real parts of the
scattering lengths indicate that there exists the attractive
interaction in the following channels: ${ \pi\Lambda_c}^{(1)}$,
${ \pi\Xi_c}^{(1/2)}$,
${ K\Xi_c}^{(0)}$,
${ K\Xi_c}^{(1)}$,
${ \bar K\Lambda_c}^{(1/2)}$,
${ \bar K\Xi_c}^{(0)}$,
${ \eta\Lambda_c}^{(0)}$,
${ \eta\Xi_c}^{(1/2)}$ , ${ \pi\Xi^\prime_c}^{(1/2)}$, ${
\pi\Sigma_c}^{(0)}$, ${ \pi\Sigma_c}^{(1)}$, ${
K\Omega_c}^{(1/2)}$, ${ K\Xi^\prime_c}^{(0)}$, ${
K\Xi^\prime_c}^{(1)}$, ${ K\Sigma_c}^{(1/2)}$, ${ \bar
K\Xi^\prime_c}^{(0)}$, ${ \bar K\Sigma_c}^{(1/2)}$, ${ \bar
K\Sigma_c}^{(3/2)}$, ${ \eta\Omega_c}^{(0)}$, ${
\eta\Xi^\prime_c}^{(1/2)}$, ${ \eta\Sigma_c}^{(1)}$, ${
\pi\Xi^*_c}^{(1/2)}$, ${ \pi\Sigma^*_c}^{(0)}$, ${
\pi\Sigma^*_c}^{(1)}$, ${ K\Omega^*_c}^{(1/2)}$, ${
K\Xi^*_c}^{(0)}$, ${ K\Xi^*_c}^{(1)}$, ${ K\Sigma^*_c}^{(1/2)}$,
${ \bar K\Xi^*_c}^{(0)}$, ${ \bar K\Sigma^*_c}^{(1/2)}$, ${ \bar
K\Sigma^*_c}^{(3/2)}$, ${ \eta\Omega^*_c}^{(0)}$, ${
\eta\Xi^*_c}^{(1/2)}$, and ${ \eta\Sigma^*_c}^{(1)}$, where the
superscripts refer to the isospin.

\begin{table}[p]
\caption{The $\phi$-$B_{\bar3}$ threshold $T$-matrices order by
order in units of fm with $\alpha'=0\pm 1$.}\label{TabTPhiB3}
\begin{tabular}{ccccccc}
\hline
~                                 &$O(\epsilon^1)$ &$O(\epsilon^2)$   &$O(\epsilon^3)$        &Total                         &Scattering Length                 &Scattering Length                  \\
                                  &                &                  &                       &                              &                                  &(Nonanalytic Approximation)        \\
\hline
$T^{(1)}_{ \pi\Lambda_c}$	&	0	&	0.65	&	-0.11	&	0.54	&	0.041	&	0.032	\\
$T^{(1/2)}_{ \pi\Xi_c}$	&	3.2	&	0.59  $\pm$  0.069 	&	0.15	&	4.  $\pm$  0.069 	&	0.3  $\pm$  0.0052 	&	0.29  $\pm$  0.0052 	\\
$T^{(3/2)}_{ \pi\Xi_c}$	&	-1.6	&	0.59  $\pm$  0.069 	&	-0.64	&	-1.7  $\pm$  0.069 	&	-0.13  $\pm$  0.0052 	&	-0.12  $\pm$  0.0052 	\\
\hline
$T^{(1/2)}_{ K\Lambda_c}$	&	-3.8	&	4.9  $\pm$  0.57 	&	-3.7	&	-2.6  $\pm$  0.57 	&	-0.17  $\pm$  0.038 	&	-0.079  $\pm$  0.038 	\\
$T^{(0)}_{ K\Xi_c}$	&	7.6	&	4.4  $\pm$  1.1 	&	0.78	&	13.  $\pm$  1.1 	&	0.85  $\pm$  0.076 	&	0.71  $\pm$  0.076 	\\
$T^{(1)}_{ K\Xi_c}$	&	0	&	5.4	&	4.1+2.8 $i$	&	9.6 + 2.8 $i$	&	0.63 + 0.18 $i$	&	0.59 + 0.18 $i$	\\
$T^{(1/2)}_{ \bar K\Lambda_c}$	&	3.8	&	4.9  $\pm$  0.57 	&	1.3+4.2 $i$	&	(10. + 4.2 $i$)  $\pm$  0.57 	&	(0.66 + 0.27 $i$)  $\pm$  0.038 	&	(0.56 + 0.27 $i$)  $\pm$  0.038 	\\
$T^{(0)}_{ \bar K\Xi_c}$	&	3.8	&	6.  $\mp$  0.57 	&	14	&	24.  $\mp$  0.57 	&	1.6  $\mp$  0.038 	&	1.4  $\mp$  0.038 	\\
$T^{(1)}_{ \bar K\Xi_c}$	&	-3.8	&	4.9  $\pm$  0.57 	&	-3.6	&	-2.5  $\pm$  0.57 	&	-0.17  $\pm$  0.038 	&	-0.075  $\pm$  0.038 	\\
\hline
$T^{(0)}_{ \eta\Lambda_c}$	&	0	&	2.1  $\pm$  0.69 	&	1.5+3. $i$	&	(3.6 + 3. $i$)  $\pm$  0.69 	&	(0.23 + 0.19 $i$)  $\pm$  0.044 	&	(0.22 + 0.19 $i$)  $\pm$  0.044 	\\
$T^{(1/2)}_{ \eta\Xi_c}$	&	0	&	5.9  $\pm$  0.17 	&	0.55+1.5 $i$	&	(6.4 + 1.5 $i$)  $\pm$  0.17 	&	(0.42 + 0.098 $i$)  $\pm$  0.011 	&	(0.41 + 0.098 $i$)  $\pm$  0.011 	\\
\hline
\end{tabular}
\end{table}

\begin{table}[p]
\caption{The $\phi$-$B_{6}$ threshold $T$-matrices order by order
in units of fm with $\alpha'=0\pm 1$.}\label{TabTPhiB6}
\begin{tabular}{ccccccc}
\hline
~                                 &$O(\epsilon^1)$ &$O(\epsilon^2)$   &$O(\epsilon^3)$        &Total                         &Scattering Length                 &Scattering Length                  \\
                                  &                &                  &                       &                              &                                  &(Nonanalytic Approximation)        \\
\hline
$T^{(1)}_{ \pi\Omega_c}$	&	0	&	0.51  $\pm$  0.42 	&	-1.3	&	-0.77  $\pm$  0.42 	&	-0.058  $\pm$  0.031 	&	-0.058  $\pm$  0.031 	\\
$T^{(1/2)}_{ \pi\Xi^\prime_c}$	&	3.2	&	0.7  $\pm$  0.21 	&	-0.78	&	3.1  $\pm$  0.21 	&	0.24  $\pm$  0.016 	&	0.23  $\pm$  0.016 	\\
$T^{(3/2)}_{ \pi\Xi^\prime_c}$	&	-1.6	&	0.7  $\pm$  0.21 	&	-1.6	&	-2.5  $\pm$  0.21 	&	-0.19  $\pm$  0.016 	&	-0.18  $\pm$  0.016 	\\
$T^{(0)}_{ \pi\Sigma_c}$	&	6.5	&	1.5	&	1.4	&	9.3	&	0.7	&	0.67	\\
$T^{(1)}_{ \pi\Sigma_c}$	&	3.2	&	0.5	&	-0.34	&	3.4	&	0.25	&	0.25	\\
$T^{(2)}_{ \pi\Sigma_c}$	&	-3.2	&	0.89	&	-0.91	&	-3.3	&	-0.24	&	-0.23	\\
\hline
$T^{(1/2)}_{ K\Omega_c}$	&	7.6	&	7.4	&	5.8+8.3 $i$	&	21. + 8.3 $i$	&	1.4 + 0.56 $i$	&	1.2 + 0.56 $i$	\\
$T^{(0)}_{ K\Xi^\prime_c}$	&	7.6	&	5.9  $\pm$  3.5 	&	9.9+8.3 $i$	&	(23. + 8.3 $i$)  $\pm$  3.5 	&	(1.6 + 0.56 $i$)  $\pm$  0.23 	&	(1.4 + 0.56 $i$)  $\pm$  0.23 	\\
$T^{(1)}_{ K\Xi^\prime_c}$	&	0	&	5.8	&	-4.3+5.6 $i$	&	1.5 + 5.6 $i$	&	0.1 + 0.37 $i$	&	0.11 + 0.37 $i$	\\
$T^{(1/2)}_{ K\Sigma_c}$	&	3.8	&	2.7  $\pm$  5.2 	&	3.3	&	9.8  $\pm$  5.2 	&	0.65  $\pm$  0.35 	&	0.56  $\pm$  0.35 	\\
$T^{(3/2)}_{ K\Sigma_c}$	&	-7.6	&	7.4	&	-6	&	-6.2	&	-0.41	&	-0.23	\\
$T^{(1/2)}_{ \bar K\Omega_c}$	&	-7.6	&	7.4	&	-4.2	&	-4.4	&	-0.3	&	-0.11	\\
$T^{(0)}_{ \bar K\Xi^\prime_c}$	&	3.8	&	5.7  $\mp$  1.7 	&	-3.1	&	6.5  $\mp$  1.7 	&	0.43  $\mp$  0.12 	&	0.36  $\mp$  0.12 	\\
$T^{(1)}_{ \bar K\Xi^\prime_c}$	&	-3.8	&	5.8  $\pm$  1.7 	&	-3.3+5.6 $i$	&	(-1.2 + 5.6 $i$)  $\pm$  1.7 	&	(-0.083 + 0.37 $i$)  $\pm$  0.12 	&	(0.0088 + 0.37 $i$)  $\pm$  0.12 	\\
$T^{(1/2)}_{ \bar K\Sigma_c}$	&	11	&	9.  $\mp$  1.7 	&	10.+1.4 $i$	&	(31. + 1.4 $i$)  $\mp$  1.7 	&	(2. + 0.092 $i$)  $\mp$  0.12 	&	(1.8 + 0.092 $i$)  $\mp$  0.12 	\\
$T^{(3/2)}_{ \bar K\Sigma_c}$	&	0	&	4.3  $\pm$  3.5 	&	-2.+5.6 $i$	&	(2.3 + 5.6 $i$)  $\pm$  3.5 	&	(0.15 + 0.37 $i$)  $\pm$  0.23 	&	(0.15 + 0.37 $i$)  $\pm$  0.23 	\\
\hline
$T^{(0)}_{ \eta\Omega_c}$	&	0	&	9.9  $\mp$  1.1 	&	-1.1+6.1 $i$	&	(8.8 + 6.1 $i$)  $\mp$  1.1 	&	(0.58 + 0.4 $i$)  $\mp$  0.069 	&	(0.59 + 0.4 $i$)  $\mp$  0.069 	\\
$T^{(1/2)}_{ \eta\Xi^\prime_c}$	&	0	&	8.3  $\pm$  0.53 	&	1.5+7.6 $i$	&	(9.8 + 7.6 $i$)  $\pm$  0.53 	&	(0.64 + 0.49 $i$)  $\pm$  0.034 	&	(0.63 + 0.49 $i$)  $\pm$  0.034 	\\
$T^{(1)}_{ \eta\Sigma_c}$	&	0	&	2.2  $\pm$  2.1 	&	0.43+3. $i$	&	(2.7 + 3. $i$)  $\pm$  2.1 	&	(0.17 + 0.2 $i$)  $\pm$  0.14 	&	(0.18 + 0.2 $i$)  $\pm$  0.14 	\\
\hline
\end{tabular}
\end{table}

\begin{table}[p]
\caption{The $\phi$-$B_{6}^*$ threshold $T$-matrices order by
order in units of fm with $\alpha'=0\pm 1$.}\label{TabTPhiB6s}
\begin{tabular}{ccccccc}
\hline
~                                 &$O(\epsilon^1)$ &$O(\epsilon^2)$   &$O(\epsilon^3)$        &Total                         &Scattering Length                 &Scattering Length                  \\
                                  &                &                  &                       &                              &                                  &(Nonanalytic Approximation)        \\
\hline
$T^{(1)}_{ \pi\Omega_c^*}$	&	0	&	0.51  $\pm$  0.42 	&	-1.3	&	-0.76  $\pm$  0.42 	&	-0.058  $\pm$  0.031 	&	-0.058  $\pm$  0.031 	\\
$T^{(1/2)}_{ \pi\Xi_c^*}$	&	3.2	&	0.69  $\pm$  0.21 	&	-0.79-0.035 $i$	&	(3.1 - 0.035 $i$)  $\pm$  0.21 	&	(0.24 - 0.0027 $i$)  $\pm$  0.016 	&	(0.23 - 0.0027 $i$)  $\pm$  0.016 	\\
$T^{(3/2)}_{ \pi\Xi_c^*}$	&	-1.6	&	0.69  $\pm$  0.21 	&	-1.6-0.035 $i$	&	(-2.5 - 0.035 $i$)  $\pm$  0.21 	&	(-0.19 - 0.0027 $i$)  $\pm$  0.016 	&	(-0.18 - 0.0027 $i$)  $\pm$  0.016 	\\
$T^{(0)}_{ \pi\Sigma_c^*}$	&	6.5	&	1.6	&	1.4+0.42 $i$	&	9.5 + 0.42 $i$	&	0.71 + 0.032 $i$	&	0.69 + 0.032 $i$	\\
$T^{(1)}_{ \pi\Sigma_c^*}$	&	3.2	&	0.4	&	-0.43-0.28 $i$	&	3.2 - 0.28 $i$	&	0.24 - 0.021 $i$	&	0.24 - 0.021 $i$	\\
$T^{(2)}_{ \pi\Sigma_c^*}$	&	-3.2	&	0.88	&	-0.95	&	-3.3	&	-0.25	&	-0.24	\\
\hline
$T^{(1/2)}_{ K\Omega_c^*}$	&	7.6	&	7.3	&	5.8+8.3 $i$	&	21. + 8.3 $i$	&	1.4 + 0.56 $i$	&	1.2 + 0.56 $i$	\\
$T^{(0)}_{ K\Xi_c^*}$	&	7.6	&	6.3  $\pm$  3.5 	&	10.+8.4 $i$	&	(24. + 8.4 $i$)  $\pm$  3.5 	&	(1.6 + 0.56 $i$)  $\pm$  0.23 	&	(1.4 + 0.56 $i$)  $\pm$  0.23 	\\
$T^{(1)}_{ K\Xi_c^*}$	&	0	&	5.3	&	-4.5+5.5 $i$	&	0.83 + 5.5 $i$	&	0.055 + 0.37 $i$	&	0.067 + 0.37 $i$	\\
$T^{(1/2)}_{ K\Sigma_c^*}$	&	3.8	&	2.7  $\pm$  5.2 	&	3.4	&	10.  $\pm$  5.2 	&	0.66  $\pm$  0.35 	&	0.57  $\pm$  0.35 	\\
$T^{(3/2)}_{ K\Sigma_c^*}$	&	-7.6	&	7.3	&	-6.1	&	-6.4	&	-0.42	&	-0.24	\\
$T^{(1/2)}_{ \bar K\Omega_c^*}$	&	-7.6	&	7.3	&	-4.2	&	-4.5	&	-0.31	&	-0.12	\\
$T^{(0)}_{ \bar K\Xi_c^*}$	&	3.8	&	4.9  $\mp$  1.7 	&	-3.6-0.048 $i$	&	(5.1 - 0.048 $i$)  $\mp$  1.7 	&	(0.34 - 0.0032 $i$)  $\mp$  0.12 	&	(0.27 - 0.0032 $i$)  $\mp$  0.12 	\\
$T^{(1)}_{ \bar K\Xi_c^*}$	&	-3.8	&	5.8  $\pm$  1.7 	&	-3.3+5.6 $i$	&	(-1.3 + 5.6 $i$)  $\pm$  1.7 	&	(-0.085 + 0.37 $i$)  $\pm$  0.12 	&	(0.0087 + 0.37 $i$)  $\pm$  0.12 	\\
$T^{(1/2)}_{ \bar K\Sigma_c^*}$	&	11	&	8.9  $\mp$  1.7 	&	10.+1.4 $i$	&	(30. + 1.4 $i$)  $\mp$  1.7 	&	(2. + 0.092 $i$)  $\mp$  0.12 	&	(1.7 + 0.092 $i$)  $\mp$  0.12 	\\
$T^{(3/2)}_{ \bar K\Sigma_c^*}$	&	0	&	4.3  $\pm$  3.5 	&	-1.9+5.6 $i$	&	(2.3 + 5.6 $i$)  $\pm$  3.5 	&	(0.16 + 0.37 $i$)  $\pm$  0.23 	&	(0.15 + 0.37 $i$)  $\pm$  0.23 	\\
\hline
$T^{(0)}_{ \eta\Omega_c^*}$	&	0	&	9.4  $\mp$  1.1 	&	-1.2+6.1 $i$	&	(8.2 + 6.1 $i$)  $\mp$  1.1 	&	(0.54 + 0.4 $i$)  $\mp$  0.07 	&	(0.55 + 0.4 $i$)  $\mp$  0.07 	\\
$T^{(1/2)}_{ \eta\Xi_c^*}$	&	0	&	8.4  $\pm$  0.53 	&	1.6+7.6 $i$	&	(10. + 7.6 $i$)  $\pm$  0.53 	&	(0.66 + 0.5 $i$)  $\pm$  0.034 	&	(0.65 + 0.5 $i$)  $\pm$  0.034 	\\
$T^{(1)}_{ \eta\Sigma_c^*}$	&	0	&	2.1  $\pm$  2.1 	&	0.36+3. $i$	&	(2.5 + 3. $i$)  $\pm$  2.1 	&	(0.16 + 0.2 $i$)  $\pm$  0.14 	&	(0.16 + 0.2 $i$)  $\pm$  0.14 	\\
\hline
\end{tabular}
\end{table}

There is an undetermined constant $\alpha'$ at $O(\epsilon^2)$. We
allow $\alpha'$ to vary from -1 to 1. Its contribution is small.
The variation of the scattering length is less than one fifth of
the central value in almost 40 channels among the total 49
channels.

From the tables, the leading order contribution from the chiral
connection dominates the total $T$-matrices for the most $\pi B$
channels. We regard one scattering channel as convergent when
    \begin{equation}
    \left\{ \begin{array}{cr}
            \left|T_{\phi B}\right|_{O(\epsilon^3)}<\frac12 \left|T_{\phi B}\right|_{O(\epsilon^2)}
            <\frac14 \left|T_{\phi B}\right|_{O(\epsilon^1)}
              ,
             & \qquad \left|T_{\phi B}\right|_{O(\epsilon^1)}\neq0\\
              \left|T_{\phi B}\right|_{O(\epsilon^3)} < \frac12 \left|T_{\phi B}\right|_{O(\epsilon^2)},
             & \qquad \left|T_{\phi B}\right|_{O(\epsilon^1)}=0
             \end{array} \right. .
    \end{equation}
With the above criteria there exist eleven convergent channels:
$T^{(1)}_{ \pi\Lambda_c}$, $T^{(1/2)}_{ \pi\Xi_c}$, 
 $T^{(1/2)}_{ \eta\Xi_c}$, 
 $T^{(3/2)}_{ \bar K\Sigma_c}$,
  $T^{(0)}_{ \eta\Omega_c}$, 
  $T^{(1/2)}_{
\eta\Xi'_c}$, 
$T^{(1)}_{ \eta\Sigma_c}$, 
$T^{(3/2)}_{ \bar K\Sigma_c^*}$,
 $T^{(0)}_{ \eta\Omega^*_c}$, $T^{(1/2)}_{
\eta\Xi^*_c}$, and $T^{(1)}_{ \eta\Sigma^*_c}$. The scattering
lengths of the above channels are positive. In other words, the
interaction between the pseudoscalar meson and heavy baryon is
attractive. The chiral expansion of the $K B$ channels
converges badly mainly due to the large mass of kaon.

For the eta meson scattering off the charmed baryon, the loop
diagrams in Fig. \ref{LoopDiag}(I) do not contribute to the real
part of the $T$-matrix at the threshold as can be seen from Eqs.
(\ref{TphiB3},\ref{TphiB6},\ref{PUYDef},\ref{Vdef}). Only the loop
diagrams in Fig. \ref{LoopDiag}(II) contribute to the real part of
the $T_{B\eta}$-matrix, which is helpful to the convergence in the
$\eta$ channel.

At present there is not enough experimental information on the
pseudoscalar meson and heavy baryon scattering. We are unable to
determine the low energy constants at $O(\epsilon^3)$. With the
very crude nonanalytic dominance approximation, we study the
convergence of the chiral expansion further in Appendix
\ref{secNonAly}. Under this approximation, the convergence becomes
better in the most channels, especially in ${ \bar
K\Lambda_c}^{(1/2)}$, ${ K\Omega_c}^{(1/2)}$, ${
K\Sigma_c}^{(3/2)}$, ${ \bar K\Omega_c}^{(1/2)}$, ${ \bar
K\Sigma_c}^{(1/2)}$, ${ K\Omega^*_c}^{(1/2)}$, ${
K\Sigma^*_c}^{(3/2)}$, ${ \bar K\Omega^*_c}^{(1/2)}$, and ${ \bar
K\Sigma^*_c}^{(1/2)}$.

In order to check where the large correction at $O(\epsilon^3)$
comes from, we separate the different contributions to
$T^{(I)}_{\phi B}$ at $O(\epsilon^3)$ in natural units of
$m_\phi/f_\phi^2$ in Table \ref{TabTPhiB3TriN}. We notice that the
tree contribution at $O(\epsilon^3)$ is really small since the
recoil correction should be suppressed for a heavy charmed baryon.
The inclusion of the excited charmed sextet does not suppress the
loop correction for the channels of the ground charmed baryons.

\begin{table}[p]
\caption{The $\phi$-$B$ threshold $T$-matrices at
$O(\epsilon^3)$ in natural units of
$m_\phi/f_\phi^2$.}\label{TabTPhiB3TriN}
\begin{tabular}{c|c|cccc|cc}
\hline
~                                 &Tree              &Loop I               &Loop II: triplet        &Loop II: sextet    &Loop II: excited sextet     &Loop total           &Nonanalytic part\\
\hline
$T^{(1)}_{ \pi\Lambda_c}$         &0                    &-0.15                    &0                       &0.04               &0.081                &-0.033               &-0.069         \\
$T^{(1/2)}_{ \pi\Xi_c}$           &0.00022              &0.085                    &0                       &-0.015             &-0.025               &0.045                &0.0018         \\
$T^{(3/2)}_{ \pi\Xi_c}$           &-0.0000068           &-0.16                    &0                       &-0.015             &-0.025               &-0.2                 &-0.18          \\
$T^{(1/2)}_{ K\Lambda_c}$         &-0.0007              &-0.48                    &0                       &0                  &0                    &-0.48                &-0.3           \\
$T^{(0)}_{ K\Xi_c}$               &0.0024               &0.55                     &0                       &-0.15              &-0.3                 &0.1                  &-0.18          \\
$T^{(1)}_{ K\Xi_c}$               &0                    &-0.072+0.36$i$           &0                       &0.21               &0.4                  &0.54+0.36$i$         &0.45+0.36$i$   \\
$T^{(1/2)}_{ \bar K\Lambda_c}$    &0.0012               &0.17+0.55$i$             &0                       &0                  &0                    &0.17+0.55$i$         &-0.013+0.55$i$ \\
$T^{(0)}_{ \bar K\Xi_c}$          &0.0007               &0.7                      &0                       &0.39               &0.75                 &1.8                  &1.5            \\
$T^{(1)}_{ \bar K\Xi_c}$          &-0.0007              &-0.56                    &0                       &0.03               &0.051                &-0.47                &-0.29          \\
$T^{(0)}_{ \eta\Lambda_c}$        &0.00037              &0.5$i$                   &0                       &0.085              &0.16                 &0.25+0.5$i$          &0.21+0.5$i$    \\
$T^{(1/2)}_{ \eta\Xi_c}$          &0.000092             &0.25$i$                  &0                       &0.03               &0.06                 &0.091+0.25$i$        &0.072+0.25$i$  \\
\hline
$T^{(1)}_{ \pi\Omega_c}$          &0                    &-0.31                    &0                       &-0.058             &-0.029               &-0.4                 &-0.39          \\
$T^{(1/2)}_{ \pi\Xi'_c}$          &0.0006               &-0.22                    &-0.028                  &0.0045             &0.0038               &-0.24                &-0.29          \\
$T^{(3/2)}_{ \pi\Xi'_c}$          &0.000075             &-0.47                    &-0.028                  &0.0045             &0.0038               &-0.49                &-0.46          \\
$T^{(0)}_{ \pi\Sigma_c}$          &0.00052              &0.17                     &0.058                   &0.12               &0.083                &0.43                 &0.3            \\
$T^{(1)}_{ \pi\Sigma_c}$          &0.0013               &0.0084                   &-0.039                  &-0.047             &-0.03                &-0.11                &-0.13          \\
$T^{(2)}_{ \pi\Sigma_c}$          &0.00015              &-0.32                    &0                       &0.018              &0.015                &-0.28                &-0.24          \\
$T^{(1/2)}_{ K\Omega_c}$          &0.0062               &0.34+1.1$i$              &0                       &0.28               &0.14                 &0.75+1.1$i$          &0.38+1.1$i$    \\
$T^{(0)}_{ K\Xi'_c}$              &0.0036               &1.2+1.1$i$               &-0.079                  &0.13               &0.067                &1.3+1.1$i$           &0.91+1.1$i$    \\
$T^{(1)}_{ K\Xi'_c}$              &0.0026               &-0.7+0.73$i$             &0.17                    &-0.019             &-0.011               &-0.56+0.73$i$        &-0.54+0.73$i$  \\
$T^{(1/2)}_{ K\Sigma_c}$          &0.0013               &0.77                     &0                       &-0.22              &-0.12                &0.43                 &0.26           \\
$T^{(3/2)}_{ K\Sigma_c}$          &-0.0026              &-1.1                     &0                       &0.21               &0.11                 &-0.78                &-0.43          \\
$T^{(1/2)}_{ \bar K\Omega_c}$     &-0.0026              &-0.97                    &0                       &0.28               &0.14                 &-0.55                &-0.19          \\
$T^{(0)}_{ \bar K\Xi'_c}$         &0.0065               &-0.56                    &0.29                    &-0.091             &-0.05                &-0.41                &-0.55          \\
$T^{(1)}_{ \bar K\Xi'_c}$         &-0.0013              &-0.56+0.73$i$            &0.045                   &0.053              &0.028                &-0.43+0.73$i$        &-0.25+0.73$i$  \\
$T^{(1/2)}_{ \bar K\Sigma_c}$     &0.0092               &0.79+0.18$i$             &0                       &0.36               &0.19                 &1.3+0.18$i$          &0.78+0.18$i$   \\
$T^{(3/2)}_{ \bar K\Sigma_c}$     &0                    &-0.14+0.73$i$            &0                       &-0.076             &-0.042               &-0.26+0.73$i$        &-0.26+0.73$i$  \\
$T^{(0)}_{ \eta\Omega_c}$         &0.0039               &1.$i$                    &0.14                    &-0.21              &-0.11                &-0.18+1.$i$          &-0.16+1.$i$    \\
$T^{(1/2)}_{ \eta\Xi'_c}$         &0.0011               &1.3$i$                   &-0.13                   &0.25               &0.13                 &0.25+1.3$i$          &0.23+1.3$i$    \\
$T^{(1)}_{ \eta\Sigma_c}$         &0.00098              &0.5$i$                   &0.07                    &-0.00035           &0.00011              &0.07+0.5$i$          &0.079+0.5$i$   \\
\hline
$T^{(1)}_{ \pi\Omega^*_c}$        &0                    &-0.31                    &0                       &-0.014             &-0.073               &-0.4                 &-0.4           \\
$T^{(1/2)}_{ \pi\Xi^*_c}$         &0.0011               &-0.22                    &-0.03-0.011$i$          &-0.00017           &0.0056               &-0.25-0.011$i$       &-0.29-0.011$i$ \\
$T^{(3/2)}_{ \pi\Xi^*_c}$         &0.0004               &-0.47                    &-0.03-0.011$i$          &-0.00017           &0.0056               &-0.49-0.011$i$       &-0.47-0.011$i$ \\
$T^{(0)}_{ \pi\Sigma^*_c}$        &-0.0016              &0.17                     &0.12+0.13$i$            &0.008              &0.14                 &0.44+0.13$i$         &0.32+0.13$i$   \\
$T^{(1)}_{ \pi\Sigma^*_c}$        &0.0042               &0.0084                   &-0.081-0.087$i$         &-0.0064            &-0.059               &-0.14-0.087$i$       &-0.16-0.087$i$ \\
$T^{(2)}_{ \pi\Sigma^*_c}$        &0.0008               &-0.32                    &0                       &-0.00068           &0.022                &-0.29                &-0.25          \\
$T^{(1/2)}_{ K\Omega^*_c}$        &0.01                 &0.34+1.1$i$              &0                       &0.066              &0.34                 &0.75+1.1$i$          &0.39+1.1$i$    \\
$T^{(0)}_{ K\Xi^*_c}$             &0.0012               &1.2+1.1$i$               &-0.03+0.0063$i$         &0.028              &0.16                 &1.3+1.1$i$           &0.95+1.1$i$    \\
$T^{(1)}_{ K\Xi^*_c}$             &0.0089               &-0.7+0.73$i$             &0.13-0.0021$i$          &-0.0039            &-0.024               &-0.6+0.73$i$         &-0.57+0.73$i$  \\
$T^{(1/2)}_{ K\Sigma^*_c}$        &0.0006               &0.77                     &0                       &-0.048             &-0.28                &0.45                 &0.26           \\
$T^{(3/2)}_{ K\Sigma^*_c}$        &-0.0012              &-1.1                     &0                       &0.049              &0.27                 &-0.79                &-0.43          \\
$T^{(1/2)}_{ \bar K\Omega^*_c}$   &-0.0012              &-0.97                    &0                       &0.066              &0.34                 &-0.56                &-0.19          \\
$T^{(0)}_{ \bar K\Xi^*_c}$        &0.018                &-0.56                    &0.21-0.0063$i$          &-0.02              &-0.11                &-0.49-0.0063$i$      &-0.61-0.0063$i$\\
$T^{(1)}_{ \bar K\Xi^*_c}$        &-0.0006              &-0.56+0.73$i$            &0.051+0.0021$i$         &0.012              &0.067                &-0.43+0.73$i$        &-0.24+0.73$i$  \\
$T^{(1/2)}_{ \bar K\Sigma^*_c}$   &0.015                &0.79+0.18$i$             &0                       &0.081              &0.45                 &1.3+0.18$i$          &0.78+0.18$i$   \\
$T^{(3/2)}_{ \bar K\Sigma^*_c}$   &0                    &-0.14+0.73$i$            &0                       &-0.016             &-0.095               &-0.25+0.73$i$        &-0.26+0.73$i$  \\
$T^{(0)}_{ \eta\Omega^*_c}$       &0.014                &1.$i$                    &0.11                    &-0.051             &-0.26                &-0.2+1.$i$           &-0.18+1.$i$    \\
$T^{(1/2)}_{ \eta\Xi^*_c}$        &0.00084              &1.3$i$                   &-0.11+0.0013$i$         &0.057              &0.31                 &0.26+1.3$i$          &0.24+1.3$i$    \\
$T^{(1)}_{ \eta\Sigma^*_c}$       &0.0034               &0.5$i$                   &0.057+0.0017$i$         &-0.00021           &-0.00043             &0.057+0.5$i$         &0.069+0.5$i$   \\
\hline
\end{tabular}
\end{table}

It is interesting to notice that the inclusion of the $B_6^*$
intermediate states does not make the convergence better. Let's
denote the contribution of the intermediate particle $X$ to the
$T$-matrix through the axial couplings in the heavy quark symmetry
limit as $\mathcal C_X$. For the $B_{\bar3}\phi$ scattering, we
can get the following ratio from Eqs. (\ref{TphiB3},\ref{PUYDef})
\begin{equation}
\left.\frac{\mathcal C_{B_6}}{\mathcal C_{B_6^*}}\right|_{\text{for } B_{\bar3}\phi \text{ scattering}}=\frac{3 g_2^2}{2g_4^2}=\frac12.
\end{equation}
For the $B_{6}\phi$ scattering,
\begin{equation}
  \left.\frac{\mathcal C_{B_6}}{\mathcal C_{B_6^*}}\right|_{\text{for } B_{6}\phi \text{ scattering}}=\frac{3 g_1^2}{2g_3^2}=2.
\end{equation}
And for the $B_{6^*}\phi$ scattering, the ratio is
\begin{equation}
  \left.\frac{\mathcal C_{B_6}}{\mathcal C_{B_6^*}}\right|_{\text{for } B_6^*\phi \text{ scattering}}=\frac{3 g_3^2}{5g_5^2}=\frac15.
\end{equation}
One notices that $C_{B_6^*}$ is larger than $C_{B_6}$ for the
$B_{\bar3}\phi$ and $B_{6^*}\phi$ scattering, while it is smaller
than $C_{B_6}$ for the $B_{6}\phi$ scattering. The correction from
the $B_6^*$ and $B_6$ states has the same sign as required from
heavy quark symmetry. Their contribution is constructive, which
worsens the convergence.

From Tables \ref{TabTPhiB6} and \ref{TabTPhiB6s}, one notices that
the numerical value of $T_{\phi B_6}^I$ is very close to that of
the corresponding $T_{\phi B_6^*}^I$ at every order. As can be
seen in Table \ref{TabTPhiB3TriN}, the contribution of the sum of
all the loop diagrams to $T_{\phi B_6}^I$ and $T_{\phi B_6^*}^I$
is almost the same, which is the manifestation of the heavy flavor
symmetry.

Sometimes the nearby resonances or possible molecular states
in the pseudoscalar meson and heavy baryon scattering channel
might also destroy the convergence of the chiral expansion. For
example, the $1/2^-$($3/2^-$) charmed baryons couple strongly to
the Goldstone boson and $1/2^+$($3/2^+$) charmed baryons based on
a unitary baryon-meson coupled-channel model in Ref.
\cite{Romanets2012}. The convergence of the chiral expansion might
improve if the $1/2^-$ and $3/2^-$ charmed baryons are
included explicitly.

One may also wonder whether the recoil correction might spoil the
convergence. In the past several years there has some progress in
the development of the $\chi$PT in the covariant form such as the
extended-on-mass-shell renormalization scheme
\cite{Fuchs2003,Geng2008,Geng2010} and infrared regularization method
\cite{Becher1999}. It will be very interesting to compare the results
within the different schemes.

We estimate the LECs at $O(\epsilon^2)$ assuming the SU(4)
flavor symmetry and using the pseudoscalar meson and nucleon-octet
coupling constants as input. However, the SU(4) flavor symmetry is
broken in nature. The convergence of the chiral expansion might
improve if the LECs could be determined more accurately.

\section{Conclusions}\label{secCon}

In this work, we have studied the pseudoscalar meson and charmed
baryon scattering length to $O(\epsilon^3)$ with HB$\chi$PT.
The convergence of the chiral expansion of some pion and eta channels is good. Because of
the large heavy baryon mass, the recoil correction is small.

It is easy to get the $T$-matrices for the pseudoscalar meson and
bottomed baryon scattering from those in Sec. \ref{secTMat} with
the corresponding parameters replaced. The numerical results do
not change much due to the heavy quark flavor symmetry.

According to our convention, the scattering length with a positive
real part indicates there is attraction in this channel, which
provides useful information on the strong interaction between the
pseudoscalar meson and heavy flavor baryon. For example, one may
have a rough idea in which channels there may (or not) exist
loosely bound molecular states composed of a heavy flavor baryon
and a pseudoscalar meson. These systems are similar to the pionic
hydrogen. Moreover, we hope our present calculation, especially
nonanalytic parts, would be useful to the chiral extrapolation of
future simulation of the pseudoscalar meson and heavy baryon
scattering on the lattice.

\section*{ Acknowledgments}
We would like to thank Lu Jun-Xu and Geng Li-Sheng for helpful discussions.
This project was supported by the National Natural Science
Foundation of China under Grants 11075004, 11021092 and Ministry
of Science and Technology of China(2009CB825200).

\appendix
\section{Some functions and constants in the $T$-matrices} \label{secFunc}

We list the functions and constants in the $T$-matrices here,
\begin{eqnarray}
&&P_1=-\frac{3g_6^2 m_K^2 H_2(0,m_\eta,m_\pi)}{4 },\quad
  P_2=-\frac{3g_2^2 m_K^2 H_2(-\delta_2,m_\eta,m_\pi)}{4 },\quad
  P_3=-\frac{g_4^2 m_K^2 H_2(-\delta_3,m_\eta,m_\pi)}{4 },\nonumber\\&&
  U_1=-\frac{m_K^2\{3g_2^2H_2(\delta_2,m_\eta,m_\pi)+2g_4^2 H_2(\delta_3,m_\eta,m_\pi)\}}{4 },\quad
  U_2=-\frac{m_K^2\{3g_1^2H_2(0,m_\eta,m_\pi)+2g_3^2 H_2(\delta_1,m_\eta,m_\pi)\}}{4 },\nonumber\\&&
  U_3=-\frac{m_K^2\{5g_5^2H_2(0,m_\eta,m_\pi)+3g_3^2 H_2(-\delta_1,m_\eta,m_\pi)\}}{12 },\quad
  W_1(m)=\frac{-3g_2^2 H_2(\delta_2,m,m)-2g_4^2 H_2(\delta_3,m,m)}{4 },\nonumber\\&&
  W_2(m)=\frac{-3g_1^2 H_2(0,m,m)-2g_3^2 H_2(\delta_1,m,m)}{4 },\quad
  W_3(m)=\frac{-5g_5^2 H_2(0,m,m)-3g_3^2 H_2(-\delta_1,m,m)}{12 },\nonumber\\&&
  Y_1(m)=-\frac{3g_6^2 H_2(0,m,m)}{4 },\quad
  Y_2(m)=-\frac{3g_2^2 H_2(-\delta_2,m,m)}{4 },\quad
  Y_3(m)=-\frac{g_4^2 H_2(-\delta_3,m,m)}{4 },\label{PUYDef}\\&&
   V(m^2,\omega)=
   \left[-\frac{\omega^3}{2 \pi ^2 f^4}\right]
  +\frac{\omega^3 \ln \frac{|m|}{\lambda }}{\pi ^2 f^4}
  -\frac{ \omega^2}{\pi^2 f^4}
    \begin{cases}
      -\sqrt{m^2-\omega^2} \arccos\left(-\frac{\omega}{|m|} \right)& m^2\geq \omega^2 \\
      \sqrt{\omega^2-m^2} \ln \frac{\sqrt{\omega^2-m^2}-\omega}{|m|} & m^2<\omega^2, \omega<0 \\
      \sqrt{\omega^2-m^2} \left(-\ln \frac{\sqrt{\omega^2-m^2}+\omega}{|m|}+i \pi \right)& m^2<\omega^2, \omega\geq 0
   \end{cases},  \label{Vdef}
\end{eqnarray}
where
\begin{eqnarray}
  &&H_1(m^2,\omega)= \left[ \frac{6m^2\omega-5\omega^3}{72\pi^2} \right]
  +\frac{2\omega^3-3m^2\omega}{24\pi^2} \ln\frac{|m|}{\lambda}
 +\frac{m^2-\omega^2}{12\pi^2}   \left\{
           \begin{array}{ll}
             -\sqrt{m^2-\omega^2} \arccos\frac{-\omega}{|m|}                          &  m^2>\omega^2\\
             \sqrt{\omega^2-m^2} \ln \frac{-\omega+\sqrt{\omega^2-m^2}}{|m|}          &  m^2<\omega^2, \omega<0\\
             \sqrt{\omega^2-m^2} (i \pi-\ln\frac{\omega+\sqrt{\omega^2-m^2}}{|m|})    &  m^2<\omega^2, \omega>0
           \end{array}
    \right.   ,\nonumber\\&&
   H_2(\omega,m,M)=\frac{1}{f^4} \left\{
           \begin{array}{ll}
             \frac{H_1(m^2,-\omega)-H_1(M^2,-\omega)}{m^2-M^2}                         \qquad\qquad &  m^2\neq M^2\\
             \left.\frac{\partial H_1(z,-\omega)}{\partial z}\right|_{\{z\to m^2\} }  \qquad\qquad &  m^2=M^2\\
           \end{array}
    \right.\label{Hdef}
\end{eqnarray}

\section{Determination of Some LECs with SU(4) Flavor Symmetry} \label{secSUFour}

The SU(4) flavor $20'$-plet includes the nucleon octet,
$\Lambda_c$ triplet, $\Sigma_c$ sextet, and $\Xi_{cc}$ triplet,
which can be expressed by a 3-rank tensor $T_{abc}$:
\begin{equation}
  T_{cba}=-T_{abc},\quad T_{cab}=-T_{abc}+T_{acb}.\quad (a,b,c=1,2,3,4 \text{~are the flavor labels.})
\end{equation}
With the SU(4) 20'-representation and the isospin,
hypercharge, and charm of the physical particles, one obtains the
individual components
\begin{eqnarray}
 && T_{112}=\frac{1}{\sqrt2}P,\quad
  T_{122}=\frac{1}{\sqrt2}N,\quad
  T_{113}=-\frac{1}{\sqrt2}\Sigma^+,\quad
  T_{223}=\frac{1}{\sqrt2}\Sigma^-,\quad
  T_{133}=-\frac{1}{\sqrt2}\Xi^0,\quad
  T_{233}=\frac{1}{\sqrt2}\Xi^-,\quad
  \nonumber\\&&
  T_{132}=-\frac{1}{\sqrt3}\Lambda,\quad
  T_{123}=\frac{1}{2}\Sigma^0-\frac{1}{2\sqrt3}\Lambda,\quad
  T_{114}=\frac{1}{\sqrt2}\Sigma_c^{++},\quad
  T_{224}=\frac{1}{\sqrt2}\Sigma_c^0,\quad
  T_{334}=\frac{1}{\sqrt2}\Omega_c^0,\quad
  T_{144}=\frac{1}{\sqrt2}\Sigma_{cc}^{++},\quad
    \nonumber\\&&
  T_{244}=\frac{1}{\sqrt2}\Sigma_{cc}^+,\quad
  T_{344}=\frac{1}{\sqrt2}\Omega_{cc}^+,\quad
  T_{142}=\frac{1}{\sqrt3}\Lambda_{c}^+,\quad
  T_{124}=\frac12\Sigma_c^+ + \frac{1}{2\sqrt3}\Lambda_{c}^+,\quad
  T_{143}=\frac{1}{\sqrt3}\Xi_c^+,\quad
  \nonumber\\&&
  T_{134}=\frac12\Xi_c^{'+} + \frac{1}{2\sqrt3}\Xi_{c}^+,\quad
  T_{243}=\frac{1}{\sqrt3}\Xi_c^0,\quad
  T_{234}=\frac12\Xi_c^{'0} + \frac{1}{2\sqrt3}\Xi_{c}^0.
\end{eqnarray}
where we have normalized $T_{abc}$ so that the Lagrangian of
the self-energy is $\bar T^{abc} (i v\cdot D) T_{abc}$. With the
chiral symmetry, parity, C-parity, and Hermiticity, we can
construct the Lagrangian
\begin{eqnarray}
  \mathcal L^{(2)}_\text{SU(4)}&=&
            \alpha_0 \bar T T \Tr[\chi_+]
           +\alpha_1 \bar T^{abc} {\tilde \chi_+}{}_a^{~d} T_{dbc}
           +\alpha_2 \bar T^{abc} {\tilde \chi_+}{}_a^{~d} T_{dcb}
           +\alpha_3 \bar T T \Tr[v\cdot u~v\cdot u]
           +\alpha_4 \bar T^{ibc} v\cdot u_i^{~n}~v\cdot u_n^{~j} T_{jbc}\nonumber\\
           &&+\alpha_5 \bar T^{ibc} v\cdot u_i^{~n}~v\cdot u_n^{~j} T_{jcb}
           +\alpha_6 \bar T^{ija} v\cdot u_i^{~m}~v\cdot u_j^{~n} T_{mna}
           +\alpha_7 \bar T^{ija} v\cdot u_i^{~m}~v\cdot u_j^{~n} T_{nma},
\end{eqnarray}
where $u_\mu$ and $\chi_+$ are similar to those in Eq.
(\ref{uxiDef}) with the extended
$\chi^\text{ext}=\mathrm{diag}(m_\pi^2,\, m_\pi^2,\,
2m_K^2-m_\pi^2, 2m_D^2-m_\pi^2),$ and
\begin{equation}
\phi^\text{ext}=\sqrt2\left(
\begin{array}{cccc}
\frac{\pi^0}{\sqrt2}+\frac{\eta}{\sqrt6}+\frac{\eta_c}{\sqrt{12}}         &\pi^+   &K^+   &\bar D^0\\
\pi^-   &-\frac{\pi^0}{\sqrt2}+\frac{\eta}{\sqrt6}+\frac{\eta_c}{\sqrt{12}}    &K^0  &D^-\\
K^-&\overline{K}^0  &-\frac{2}{\sqrt6}\eta+\frac{\eta_c}{\sqrt{12}}  &D_s^-\\
D^0 &D^+  &D_s^+   &\frac{-3\eta_c}{\sqrt{12}}
\end{array}\right).
\end{equation}

$\mathcal L^{(2)}_\text{SU(4)}$ contains $\mathcal
L^{(2)}_{B_8\phi}$ for the nucleon octet, $\mathcal
L^{(2)}_{B_{\bar3}\phi}$ for the $\Lambda_c$ triplet, $\mathcal
L^{(2)}_{B_6\phi}$ for the $\Sigma_c$ sextet, and so on. So
comparing $\mathcal L^{(2)}_{B_8\phi}$ with that in Ref.
\cite{Liu2007} and using their values of LECs
 $b_0$, $b_D$, $b_F$, $d_0$, $d_1$, $d_D$, $d_F$, we get
\begin{eqnarray}
&&\alpha_0=b_0+\frac23 b_D=-0.79{~\rm GeV}^{-1},\quad
  \alpha_1=4b_F=-1.96{~\rm GeV}^{-1},\quad
  \alpha_2=-2b_D-2b_F=0.89{~\rm GeV}^{-1},\quad
  \alpha_3=\frac{\alpha'}{4\pi f},\nonumber\\&&
  \alpha_4=8d_0+4d_1+8d_D-4\frac{\alpha'}{4\pi f}=-4.13{~\rm GeV}^{-1}-4\frac{\alpha'}{4\pi f},\quad
  \alpha_5=-4d_1-4d_D-4d_F=-0.19{~\rm GeV}^{-1},\nonumber\\&&
  \alpha_6=8d_0+8d_D-8d_F-4\frac{\alpha'}{4\pi f}=-0.67{~\rm GeV}^{-1}-4\frac{\alpha'}{4\pi f},\nonumber\\&&
  \alpha_7=-8d_0-4d_1-8d_D+8d_F+4\frac{\alpha'}{4\pi f}=-1.00{~\rm GeV}^{-1}+4\frac{\alpha'}{4\pi f},
\end{eqnarray}
with a still unknown dimensionless constant $\alpha'$. Then
comparing $\mathcal L^{(2)}_{B_{\bar3}\phi}$ and $\mathcal
L^{(2)}_{B_6\phi}$ with Eq. (\ref{LBPhiTwo}), one gets
\begin{eqnarray}
&&\bar c_0=\frac12\alpha_0-\frac{1}{36}\alpha_1+\frac{1}{36}\alpha_2=-0.32{~\rm GeV}^{-1},\quad
  \bar c_1=\frac{5}{12} \alpha_1+\frac13 \alpha_2=-0.52{~\rm GeV}^{-1},\nonumber\\&&
  \bar c_2=2\alpha_3+\frac{5}{12}\alpha_4+\frac13 \alpha_5
   =-1.78{~\rm GeV}^{-1}+\frac13\frac{\alpha'}{4\pi f},\quad
  \bar c_3=-\alpha_3-\frac{1}{12}\alpha_6+\frac{1}{12}\alpha_7
   =-0.03{~\rm GeV}^{-1}-\frac13\frac{\alpha'}{4\pi f},\nonumber\\&&
  c_0=\alpha_0-\frac16\alpha_1-\frac16\alpha_2=-0.61{~\rm GeV}^{-1},\quad
  c_1=\frac12 \alpha_1=-0.98{~\rm GeV}^{-1},\nonumber\\&&
  c_2=\frac12 \alpha_4=-2.07{~\rm GeV}^{-1}-2\frac{\alpha'}{4\pi f},\quad
  c_3=\frac12\alpha_6+\frac12 \alpha_7=-0.84{~\rm GeV}^{-1},\quad
  c_4=\alpha_3=\frac{\alpha'}{4\pi f}.
\end{eqnarray}

\section{Nonanalytic dominance approximation and the convergence of the chiral expansion}\label{secNonAly}

The analytic terms from loop corrections are the polynomials of
$\epsilon$ possessing the symmetries of Lagrangians. They can be
absorbed by the LECs. In other words, the tree and partial loop
corrections have the same chiral structure. One may divide the
$T$-matrix into the analytic and nonanalytic part. The analytic
contribution originates from both loop and tree diagrams, while
the nonanalytic contribution originates only from the loop graphs.
One may also use the nonanalytic part to discuss the convergence
of the $T$-matrix since the LECs of the third order can not be
determined now.

For the $\pi\pi$ scattering length, the ratio of the analytic
contribution $a_{\pi\pi}^{{\rm anal},I}$ to the nonanalytic
contribution $a_{\pi\pi}^{L,I}$ is small\cite{Bijnens1996},
\begin{equation}
  \left.\frac{a_{\pi\pi}^{\rm anal,0}}{a_{\pi\pi}^{L,0}}\right|_{O(p^4)}=\frac{0.005}{0.039}, \quad
  \left.\frac{a_{\pi\pi}^{\rm anal,0}}{a_{\pi\pi}^{L,0}}\right|_{O(p^6)}=\frac{0.001}{0.016}, \quad
  \left.\frac{a_{\pi\pi}^{\rm anal,0}-a_{\pi\pi}^{\rm anal,2}}{a_{\pi\pi}^{L,0}-a_{\pi\pi}^{L,2}}\right|_{O(p^4)}
  =\frac{0.006}{0.036}, \quad
  \left.\frac{a_{\pi\pi}^{\rm anal,0}-a_{\pi\pi}^{\rm anal,2}}{a_{\pi\pi}^{L,0}-a_{\pi\pi}^{L,2}}\right|_{O(p^6)}
  =\frac{0.001}{0.015}. \nonumber
\end{equation}

Because of lack of enough data, we are still unable to estimate
the LECs at $O(\epsilon^3)$ accurately. As a very crude
approximation, we invoke the ``nonanalytic dominance
approximation'' to check the convergence of the chiral expansion,
which assumes large cancellation of the analytic terms between
loop and tree diagrams. Under this approximation, we list the
scattering lengths with the nonanalytic approximation in the last
column of Tables \ref{TabTPhiB3}, \ref{TabTPhiB6},
\ref{TabTPhiB6s}. The difference between the last two columns of
the tables could be regarded as a measure of the error resulting
from LECs at $O(\epsilon^3)$.

In our present calculation, some polynomials of $\epsilon$, such
as $m_\phi^3$, $m_\phi^2 \delta$, appear like nonanalytic in quark
mass $m_q$ at first sight. However, we have checked that the
polynomials in our results are analytic in quark mass since the
momentum of the external boson at the threshold also contributes a
factor $ m_\phi$, which is simply a kinematical factor. We can
extract the nonanalytic contribution with the new functions of $V$
and $H_1$ by dropping the analytic terms in the squared brackets in
Eqs. (\ref{Vdef}, \ref{Hdef}).

Comparing the total loop contribution with the sole nonanalytic
part in Table \ref{TabTPhiB3TriN}, we notice that the chiral
expansion does become better when appropriate LECs absorb the
analytic contribution. There are only 13 channels where the
magnitude of the loop correction is smaller than 1/4 in unit of
$m_\phi/f_\phi^2$. In contrast there are 19 channels where the
magnitude of the nonanalytic case is smaller than 1/4 in unit of
$m_\phi/f_\phi^2$. There are 9 badly convergent channels where
the magnitude of the total loop is larger than 2/3. There are only
5 badly convergent channels where the nonanalytic correction is
larger than 2/3. We have also checked our previous results for the
excited charmed meson and pseudoscalar meson scattering lengths.
The nonanalytic terms are smaller than the total loop
contributions in the most channels \cite{Liu2011a}.


\begin{thebibliography}{10}

\bibitem{Adachi2011}
I.~Adachi and et~al. (Belle~Collaboration), arXiv: 1105.4583 .

\bibitem{Sun2011}
Z.-F. Sun, \emph{et~al.}, Phys. Rev. D \textbf{84}, 054002 (2011).

\bibitem{Voloshin2004}
M.~Voloshin, Phys Lett B \textbf{579}, 316  (2004).

\bibitem{Lee2011}
N.~Lee, Z.-G. Luo, X.-L. Chen, and S.-L. Zhu, Phys. Rev. D \textbf{84}, 014031
  (2011).

\bibitem{Meguro2011}
W.~Meguro, Y.-R. Liu, and M.~Oka, Phys Lett B \textbf{704}, 547  (2011).

\bibitem{Wong2004}
C.-Y. Wong, Phys. Rev. C \textbf{69}, 055202 (2004).

\bibitem{Close2004}
F.~E. Close and P.~R. Page, Phys Lett B \textbf{578}, 119  (2004).

\bibitem{Liu2008a}
X.~Liu, Y.-R. Liu, W.-Z. Deng, and S.-L. Zhu, Phys. Rev. D \textbf{77}, 034003
  (2008).

\bibitem{Dai2008}
Y.-B. Dai, X.-Q. Li, S.-L. Zhu, and Y.-B. Zuo, Eur.Phys.J. C \textbf{55}, 249
  (2008).

\bibitem{Flynn2007}
J.~Flynn and J.~Nieves, Phys.Rev.D \textbf{75}, 074024 (2007).

\bibitem{Ohki2008}
H.~Ohki, H.~Matsufuru, and T.~Onogi, Phys. Rev. D \textbf{77}, 094509 (2008).

\bibitem{Detmold2011}
W.~Detmold, C.~J.~D. Lin, and S.~Meinel, Phys.Rev.D \textbf{84}, 094502 (2011).

\bibitem{Gong2011}
M.~Gong, \emph{et~al.}, PoS(Lattice 2010)106. [arXiv:1103.0589] .

\bibitem{Wise1992}
M.~B. Wise, Phys. Rev. D \textbf{45}, R2188 (1992).

\bibitem{Geng2010}
L.~Geng, N.~Kaiser, J.~Martin-Camalich, and W.~Weise, Phys.Rev.D \textbf{82},
  054022 (2010).

\bibitem{Liu2011a}
Z.-W. Liu, Y.-R. Liu, X.~Liu, and S.-L. Zhu, Phys. Rev. D \textbf{84}, 034002
  (2011).

\bibitem{Gamermann2007}
D.~Gamermann, E.~Oset, D.~Strottman, and M.~J.~V. Vacas, Phys. Rev. D
  \textbf{76}, 074016 (2007).

\bibitem{Guo2009}
F.~Guo, C.~Hanhart, and U.~Mei{\ss}ner, Eur.Phys.J. A \textbf{40}, 171 (2009).

\bibitem{Lutz2011}
M.~F.~M. Lutz, D.~Samart, and A.~Semke, arXiv:1107.1324 .

\bibitem{Detmold2012}
W.~Detmold, C.~J.~D. Lin, and S.~Meinel, arXiv:1203.3378 .

\bibitem{Hall2011}
J.~Hall, \emph{et~al.}, Physical Review D \textbf{84}, 114011 (2011).

\bibitem{Kaiser2001}
N.~Kaiser, Phys. Rev. C \textbf{64}, 045204 (2001).

\bibitem{Liu2007}
Y.-R. Liu and S.-L. Zhu, Phys.Rev.D \textbf{75}, 034003 (2007).

\bibitem{Gasser1985}
J.~Gasser and H.~Leutwyler, Nucl.Phys. B \textbf{250}, 465 (1985).

\bibitem{Liu2011}
Z.-W. Liu, Y.-R. Liu, and S.-L. Zhu, Phys. Rev. D \textbf{83}, 034004 (2011).

\bibitem{Bernard2006}
V.~Bernard and U.~Mei{\ss}ner, Phys Lett B \textbf{639}, 278 (2006).

\bibitem{procura2007}
M.~Procura, B.~Musch, T.~Hemmert, and W.~Weise, Phys.Rev.D \textbf{75}, 014503
  (2007).

\bibitem{PDG2006}
W.-M. Yao and et~al., J. Phys. G \textbf{33}, 1 (2006).

\bibitem{Liu2007a}
Y.-R. Liu and S.-L. Zhu, Eur.Phys.J.C \textbf{52}, 177 (2007).

\bibitem{Escribano2005}
R.~Escribano and J.~Fr{\`e}re, JHEP \textbf{0506}, 029 (2005).

\bibitem{Romanets2012}
O.~Romanets, \emph{et~al.}, Phys. Rev. D \textbf{85}, 114032 (2012).

\bibitem{Fuchs2003}
T.~Fuchs, J.~Gegelia, G.~Japaridze, and S.~Scherer, Phys. Rev. D \textbf{68},
  056005 (2003).

\bibitem{Geng2008}
L.~Geng, J.~Camalich, L.~Alvarez-Ruso, and M.~Vacas, Phys Rev Lett
  \textbf{101}, 222002 (2008).

\bibitem{Becher1999}
T.~Becher and H.~Leutwyler, Eur.Phys.J. C \textbf{9}, 643 (1999).

\bibitem{Bijnens1996}
J.~Bijnens, \emph{et~al.}, Phys Lett B \textbf{374}, 210 (1996).

\end{thebibliography}
\end{document}